\documentclass[useAMS,usenatbib]{mnras}
\usepackage{amsmath,psfig,epsfig,graphics}
\usepackage{graphicx}
\usepackage{epstopdf}
\usepackage{float}
\usepackage{enumerate}
\usepackage{natbib}
\usepackage{threeparttable}
\usepackage[normalem]{ulem}

\def\chandra    {{\em Chandra}\/}

\def\hst   {{\em HST}\/}
\def\xmm        {XMM-{\em Newton}\/}
\def\rosat      {{\em ROSAT}\/}

\def\vla        {{\em VLA}\/}
\def\gmrt       {{\em GMRT}\/}
\def\arcdeg{\hbox{$^\circ$}}
\def\arcmin{\hbox{$^\prime$}}
\def\arcsec{\hbox{$^{\prime\prime}$}}

\newcommand{\RNum}[1]{\uppercase\expandafter{\romannumeral #1\relax}}

\begin{document}

\title[AGN Feedback in Galaxy Group 3C~88]{AGN Feedback in Galaxy Group 3C~88: Cavities, Shock and Jet Reorientation}

\author[Liu et al.]
{Wenhao Liu${}^1$\thanks{E-mail: wl0014@uah.edu}, Ming Sun${}^1$\thanks{E-mail: ms0071@uah.edu}, Paul Nulsen${}^{2,3}$, Tracy Clarke${}^4$, Craig Sarazin${}^5$,\newauthor William Forman${}^2$, Massimo Gaspari${}^6$\thanks{Einstein and Spitzer Fellow}, Simona Giacintucci${}^4$, Dharam Vir Lal${}^7$, \newauthor Tim Edge${}^1$\\
\\
$^{1}$ Department of Physics and Astronomy, University of Alabama in Huntsville, Huntsville, AL 35899, USA\\
${}^2$ Harvard-Smithsonian Center for Astrophysics, 60 Garden Street, Cambridge, MA 02138, USA \\
${}^3$ ICRAR, University of Western Australia, 35 Stirling Hwy, Crawley, WA 6009, Australia\\
${}^4$ Naval Research Laboratory, Washington, DC 20375, USA \\
${}^5$ Department of Astronomy, University of Virginia, Charlottesville, VA 22904, USA\\
${}^6$ Department of Astrophysical Sciences, Princeton University, Princeton, NJ 08544, USA\\
${}^7$ National Centre for Radio Astrophysics, Pune University Campus, Ganeshkhind, Pune 411 007, India
}

\pagerange{\pageref{firstpage}--\pageref{lastpage}} \pubyear{2018}

\maketitle

\label{firstpage}

\begin{abstract}
	We present results from the deep \chandra\ observation (105 ksec), together with new Giant Metrewave Radio Telescope and Very Large Array data of the AGN outburst in the radio-loud galaxy group 3C~88. The system shows a prominent X-ray cavity on the eastern side with a diameter of $\sim50$ kpc at $\sim28$ kpc from the nucleus. The total enthalpy of the cavity is $3.8\times10^{58}$ erg and the average power required to inflate the X-ray bubble is $\sim2.0\times10^{43}$ erg s$^{-1}$. From surface brightness profiles we detect a shock with a Mach number of $M=1.4\pm0.2$, consistent with the value obtained from temperature jump. The shock energy is estimated to be $1.9\times10^{59}$ erg. The size and total enthalpy of the cavity in 3C~88 are the largest known in galaxy groups, as well as the shock energy. The eastern X-ray cavity is not aligned with the radio jet axis. This factor, combined with the radio morphology, strongly suggests jet reorientation in the last tens of million years. 
	The bright rim and arm features surrounding the cavity show metallicity enhancement, suggesting they originated as high metallicity gas from the group center, lifted by the rising X-ray bubbles. 
	Our \chandra\ study of 3C~88 also reveals that galaxy groups with powerful radio AGN can have high cavity power, although deep X-ray observations are typically required 
	to confirm the cavities in galaxy groups.
\end{abstract}


\section{Introduction}
\label{intro}
The X-ray observations from \chandra\ and \xmm\ have revealed that the radiative cooling time of the X-ray emitting
gas at the centers of many groups and clusters is less than 1 Gyr \citep[e.g.,][]{OSullivan17}. In the absence of heating, 
a cooling flow occurs and the hot gas cools, condenses, and flows toward the center \citep[e.g.,][]{Fabian94}. 
However, much less cool gas, below 0.5 - 1 keV, has been found by \chandra\ and \xmm\ than predicted by the cooling flow models
\citep[e.g.,][]{David01, Peterson01, Fabian06}, suggesting there is a heating source which compensates for the radiative cooling. 	
Among many proposed possibilities, the feedback from the central active galactic nucleus (AGN) 
appears to be the most promising one \citep[for a review see][]{McNamara07}.

There is clear observational evidence of AGN heating. The brightest central galaxies (BCGs) of clusters and groups
are more likely to host a radio-loud AGN \citep[e.g.,][]{Burns90, Best07, Sun09b, Mittal09}. High-resolution X-ray observations
have revealed disturbed structures, such as shocks and cavities, in the cores of clusters, groups, and elliptical galaxies
\citep[e.g.,][]{Birzan04,Birzan08, Nulsen05a,Nulsen05b, Forman05, Dunn06a, Fabian06, Croston08, Croston11,
Baldi09, Randall15, Kraft12, Su17}.
These features are caused by the radio AGN jets. As the radio jets extend outwards and inflate radio lobes, the
X-ray emitting gas is pushed aside, creating cavities (bubbles) visible in the X-ray images \citep{Churazov01}. The total enthalpy 
of the buoyantly rising cavities is found to be sufficient to offset the radiative cooling \citep{Birzan04, Rafferty06,
Nulsen07, Hlavacek-Larrondo12}. Meanwhile, weak ``cocoon'' shocks, as expected from models of jet-fed radio lobes 
\citep{Scheuer74,Heinz98}, provide additional heating from AGN outbursts, although they are difficult to
detect and only a few examples have been found 
\citep[e.g.,][]{Nulsen05a,Nulsen05b,Forman05,Baldi09,Gitti10,Croston11,Randall15}. However, the details of the AGN
feedback process, e.g., how the energy from AGN outburst is deposited in the ambient ICM and the accretion process, are poorly understood.

Ever since the discovery of X-ray cavities in cool cores, they have been an active topic observationally and theoretically.
However, AGN heating has not been well studied in galaxy groups \citep{McNamara07, Gitti12}, compared
to numerous detailed studies in clusters \citep[see also][]{Sun12, McNamara12}.
Groups are gas-poor at their centers and group cool cores are smaller than cluster cool cores \citep[e.g., see Fig. 5 in][]{Sun12}.
High-resolution 3D hydrodynamic simulations in massive groups show that self-regulated mechanical AGN feedback is key to prevent severe overcooling and overheating \citep[e.g.,][]{Gaspari11}. The heating occurs through bubble inflation, cocoon shocks, and later turbulent mixing \citep{Churazov01,McNamara07,Soker16}. The feedback power is directly linked to the X-ray luminosity: the hot plasma halo is the progenitor reservoir out of which multiphase gas condenses and later boosts the supermassive black hole (SMBH) accretion via a process known as chaotic cold accretion (CCA) \citep{Gaspari17}. 
Hot Bondi accretion has been shown by many works that it generally produces too low power and has a poor self-regulation disconnected from the X-ray cool core \citep[e.g.,][]{Gaspari11,Gaspari12a,Gaspari17a,Gaspari17,Gaspari18, Yang16, Barai16, Voit17}.
On the other hand, galaxy groups provide an excellent opportunity to study AGN feedback. 
The impact of AGN outbursts is much more pronounced in low-mass systems due to their shallow gravitational potentials \citep[e.g.,][]{Giodini10}.
The imprint of non-gravitational processes (including AGN feedback) is observed in the scaling relations of e.g.,
gas fraction and entropy \citep[e.g.,][]{Sun09}.
In addition, the outburst interval, typically of the order of the central cooling time which is much shorter in groups 
than in clusters \citep[e,g.,][]{OSullivan17}, is expected to be more frequent and gentle, which is essential to properly solve
the cooling flow problem \citep[e.g.,][]{Gaspari11}. Thus, we expect more frequently to detect multiple feedback imprints 
in the same group, e.g., NGC~5813 \citep{Randall15} and NGC~5044 \citep{David17}.
It is, therefore, crucial to build a census of AGN feedback imprints in galaxy groups, which is currently lacking, to understand
the evolution of the hot intragroup medium, and the hosted SMBH.

Here, we present a detailed study of a nearby galaxy group, with deep \chandra\ observations, to explore AGN feedback.
3C~88 (UGC~02748) is the brightest central galaxy of a galaxy group at $z=0.0302$, which is
included in the 400 square degree \rosat\ PSPC galaxy cluster survey catalog \citep{Burenin07}.
It has been studied at many wavelengths. Optically, it is classified as a low-ionization nuclear emission-line region \citep[LINER;][]{Lewis03}, 
containing a low-power AGN with the 2-10 keV luminosity of $\sim 5\times10^{41}$ erg s$^{-1}$ \citep{Gliozzi08b}. 
The \hst\ image of 3C~88 shows a faint dust lane within the central $\sim$ 0.4 kpc radius, roughly perpendicular to the radio jet \citep{deKoff00}. There is H$\alpha$+[N II] emission detected from an elliptical region roughly centered on the galaxy nucleus with an extent of $\sim5\arcsec$ \citep{Baum88}.
In the radio band, the FR classification of 3C~88 is arguable. Based on its morphology, 3C~88 is classified as an FR II radio 
galaxy \citep[e.g.,][]{Martel99,Marchesini04}, while based on its radio power it is considered as the transition source between
FR I and FR II \citep[e.g.,][]{Baum92,Gliozzi08b}.
The radio images of 3C~88, from Giant Metrewave Radio Telescope (\gmrt) at 610 MHz and Very Large Array (\vla) at 1.4 GHz and 4.9 GHz, are shown in Fig. 1. 3C~88 shows a prominent radio core and jet to the northeast. Radio lobes extend from the northeast to the east and from the southwest to the west, which makes 3C~88 appear like a Z-shaped radio source. 
Some radio features are marked in Fig. 1.
There are four ``hotspot''-like features near the edge of the radio source. However, three of them are not aligned with the prominent northeastern jet. Two of them are indeed aligned, but $\sim19\arcdeg$ from the jet.
The bolometric radio luminosity is ~$5.48\times10^{41}$ erg s$^{-1}$ from 10 MHz to 5 GHz, which
is close to the FR I/II boundary line.
3C~88's radio morphology strongly suggests a jet reorientation from east-west to northeast-southwest.
As revealed by the shallower \chandra\ observations \citep{Sun09b,Zou16}, 3C~88 is a cool core group with an average 
temperature of $\sim0.9$ keV, and a total bolometric X-ray luminosity of $3.6\times10^{42}$ erg s$^{-1}$.

This paper is organized as follows. The \chandra\ and radio observations and data reduction are presented in Section 2. 
The spatial and spectral analysis are in Section 3. In Section 4 the properties of cavities 
are discussed. In Section 5 we discussed the warm gas in 3C~88. Section 6 and  Section 7 
contains the discussion and the summary of the paper. 
We assumed the cosmology with $H_{\rm 0}=70$ km s$^{-1}$ Mpc$^{-1}$, $\Omega_{\rm M}=0.3$,
and $\Omega_{\rm \Lambda}=0.7$. At a redshift of $z=0.0302$, the angular diameter distance is 124.7 Mpc and 1\arcsec\ corresponds to 0.604 kpc.
All error bars are quoted at $1\sigma$ confidence level, unless otherwise specified.

\section{\chandra\ and Radio Observations and Data Analysis}
\label{DataAnalysis}
\subsection{\chandra\ observations}
\begin{table}
	\protect\caption{Summary of \chandra\ Observations of 3C~88 (PI: Sun)}
\begin{tabular}{|c|c|c|c|}
\hline 
ObsID & Date Obs & Total Exp    & Cleaned Exp \\
      &          & (kses)         & (ksec)        \tabularnewline
\hline 
11977 & 2009 Oct 06 & 49.62 & 49.13 \tabularnewline
11751 & 2009 Oct 14 & 19.92 & 19.44 \tabularnewline
12007 & 2009 Oct 15 & 34.62 & 33.77 \tabularnewline
\hline 
\end{tabular}
\label{tab_obs}
\end{table}
3C~88 was observed by \chandra\ for $\sim105$ ksec split into three observations 
(Obs. IDs 11751, 11977, and 12007, PI: Sun) in October 2009 with the 
Advanced CCD Imaging Spectrometer (ACIS). All three observations (aimpoint on chip S3) were taken in 
Very Faint (VFAINT) mode and centered on the cool core. The details of the \chandra\ observations are 
summarized in Table \ref{tab_obs}. There was one ACIS-I observation taken in 2008 with a short exposure of
$\sim11$ ksec \citep{Sun12,Zou16}. In this study we focus on the new, deep data with ACIS-S.
The data were analyzed using the CIAO 4.9 and CALDB version 4.7.4 from \chandra\ X-ray Center. 
For each observation the level 1 event files were reprocessed using
{\tt CHANDRA\_REPRO} script to account for afterglows, bad pixels, 
charge transfer inefficiency, and time-dependent gain correction. The improved background filtering 
was also applied by setting {\tt CHECK\_VF\_PHA} as yes to remove bad events that are likely 
associated with cosmic rays. The background light curve extracted from a source-free region was 
filtered with {\tt LC\_CLEAN} script\footnote{http://asc.harvard.edu/contrib/maxim/acisbg/} to identify any period affected by background flares.
There were no strong background flares for all observations and the resulting cleaned exposure time is shown in Table \ref{tab_obs}.

The point sources were detected in the combined 0.5-7.0 keV count image using the CIAO tool {\tt WAVDETECT}, with the variation of the point spread function across the field considered.
The detection threshold was set to $10^{-6}$ and the scales are from 1 to 16 pixels, increasing in
steps of a factor of $\sqrt{2}$. All detected sources were visually inspected and masked in the analysis.
The weighted exposure map was generated to account for quantum efficiency, vignetting
and the energy dependence of the effective area assuming an absorbed APEC model with $kT = 1.0$ keV,
$N_{\textrm{H}} = 1.12\times10^{21}$ cm$^{-2}$, and abundance of 0.3 $Z_{\odot}$ at the redshift 
$z = 0.0302$ in the 0.7-2.0 keV.
The instrumental background was estimated using the stowed background, which contains only the particle background. 
The blank sky files contain both the particle background and
the cosmic X-ray background. Rescaling blank sky background to remove
the particle background may over- or under-estimate the cosmic X-ray
background, and would generally require a double-subtraction. While the stowed
background includes only the particle background, we can remove it and
then model local X-ray background. In our studies we preferred the stowed background. 
For each observation the standard stowed file for each chip was reprojected to match the time dependent aspect
solution, and normalized to match the count rate in the 9.5-12.0 keV band.

Proper background modeling is essential for spectral analysis, especially in low surface brightness regions. 
Since the diffuse emission from the galaxy group 3C~88 does not extend over the entire field of view,
the region beyond $\sim8\arcmin$ from 3C~88, where the surface brightness is approximately constant as shown in Fig. \ref{sbbkg},
can be used to model the local X-ray background.
For each observation we extracted the spectrum
from the region ($>9\arcmin$) from ACIS-S1, ACIS-S2, and ACIS-I chips separately (I2 and I3 chips are 
also active for our ACIS-S observations), since the CCD response varies for different chips. All 9 spectra
were fitted simultaneously with two thermal components at solar abundance and zero redshift plus an absorbed
power-law with an index of 1.5 to determine the ``local'' X-ray background from this region in our observations. One thermal component is unabsorbed with the
temperature fixed at 0.1 keV to account for the emission from the Local Hot Bubble \citep{Snowden98,Sun09,Liu17},
the other thermal component is absorbed with the temperature allowed to vary to account for the emission from
the Galactic halo. The absorption model is {\tt TBABS} with a fixed column 
density of $N_{\textrm{H}} = 1.12\times10^{21}$ cm$^{-2}$. All normalizations are allowed to vary. The best fitting model gives
the Galactic halo temperature of $kT = 0.20^{+0.01}_{-0.02}$ keV. We also compared the derived soft X-ray background
flux with the ROSAT All-Sky Survey (RASS) R45 values around the galaxy group \citep[see details in ][]{Sun09}. 
We found the observed soft X-ray background flux of $2.94^{+0.37}_{-0.34}\times10^{-12}$ 
erg s$^{-1}$ cm$^{-2}$ deg$^{-2}$ in the 0.47-1.21 keV band and R45 value of $113.3\pm8.0$, 
which is consistent with the relation in Fig. 2 of \citet{Sun09}.
For spectral fitting, we used XSPEC version 12.9.1 and AtomDB 3.0.8, assuming the solar
abundance table by \citet{aspl}.

We also examined the X-ray absorption column density towards 3C~88. The weighted column density of the total Galactic
hydrogen is $1.12\times10^{21}$ cm$^{-2}$ (vs. the weighted HI column density of $0.79\times10^{21}$ cm$^{-2}$), 
calculated by the ``NHtot'' tool\footnote{http://www.swift.ac.uk/analysis/nhtot/index.php} including both the atomic hydrogen 
and the molecular hydrogen column density \citep{Willingale13}. We fitted the spectra extracted from a circular region of
radius 6\arcmin\ centered on the central AGN for each observation (the central 1.5\arcmin\ was excluded). The model
includes an absorbed thermal component for the diffuse emission from the group plus the diffuse X-ray 
background component. The background was modeled and re-scaled based on the local X-ray background estimation. 
The absorption parameter, the temperature,
abundance, and the normalization of the source component were allowed to vary. We obtained the best fitting value of
$N_{\textrm{H}} = 1.21\pm0.16\times10^{21}$ cm$^{-2}$, which is consistent with the weighted total hydrogen column density.
We found that our results are not very sensitive to the value of the column density, e.g., the uncertainties of
the temperature and the Mach number are generally within 5\%, and in some cases $\sim10$\%.

\subsection{Radio Observations}
3C~88 was observed with the \gmrt\ at 610 MHz on 2010 May 7 (project 18\_009, PI: Sun). The data were collected using the software correlator and default spectral-line observing mode. The data sets were calibrated and reduced using the NRAO Astronomical Image Processing System package (AIPS) as described in \citet{Giacintucci11}. Self-calibration was applied to reduce residual phase variations and improve the quality of the final images. Due to the large field of view of the \gmrt, we used the wide-field imaging technique at each step of the phase self-calibration process to account for the non-planar nature of the sky. The final image was produced using the multi-scale CLEAN implemented in the AIPS task IMAGR, which results in better imaging of extended sources compared to the traditional CLEAN \citep[e.g.,][]{Greisen09}. The final image at 610 MHz has a resolution of 7.1\arcsec$\times$5.4\arcsec\ and an rms noise level (1$\sigma$) of 100 $\mu$Jy/beam (Fig. \ref{radio}). A total flux of $8.8\pm0.7$ Jy is measured for 3C~88 in this image.

NRAO \vla\ observations of 3C~88 were obtained on 2011 April 15 as part of program SB0517 (PI: Sun). Data presented in this paper were taken in the \vla\ B configuration with two tunings of bandwidth 128 MHz each centered on frequencies of 1327 and 1455 MHz. The data were observed in spectral-line mode. The observations were calibrated within a modified Common Astronomy Software Applications (CASA) pipeline in version 5.1.2-4 of CASA following standard  calibration procedures. Minor modifications to the pipeline were required to enable it to accurately recognize observing intents for calibrators. The flux scale was set using 3C~48. The imaging was undertaken using the multi-scale cleaning to accurately recover emission on scales relevant to the source structure. Imaging algorithms also included both multifrequency synthesis with two Taylor terms to more accurately represent the spectral dependence of the emission and w-projection to enable wide-field imaging corrections for non-coplanar effects. The image presented in Fig. \ref{radio} has a resolution of 3.9\arcsec$\times$3.7\arcsec\ and an rms level of 45 $\mu$Jy/beam (1$\sigma$).

We also show the archival \vla\ image at 4.9 GHz in Fig. \ref{radio}. The observation was taken on Apr. 22, 1984 with the C configuration. This image was produced as part of the NRAO \vla\ Archive Survey, (c) AUI/NRAO.\footnote{The NVAS can [currently] be browsed through http://archive.nrao.edu/nvas/} 
The beam size is 4.30\arcsec\ and the rms level is 112 $\mu$Jy/beam (1$\sigma$).

\section{Spatial and Spectral Analysis}
\subsection{Image Analysis and Radial Profiles}
\label{SBshock}
\begin{figure*}
	\centerline{\includegraphics[scale=0.31]{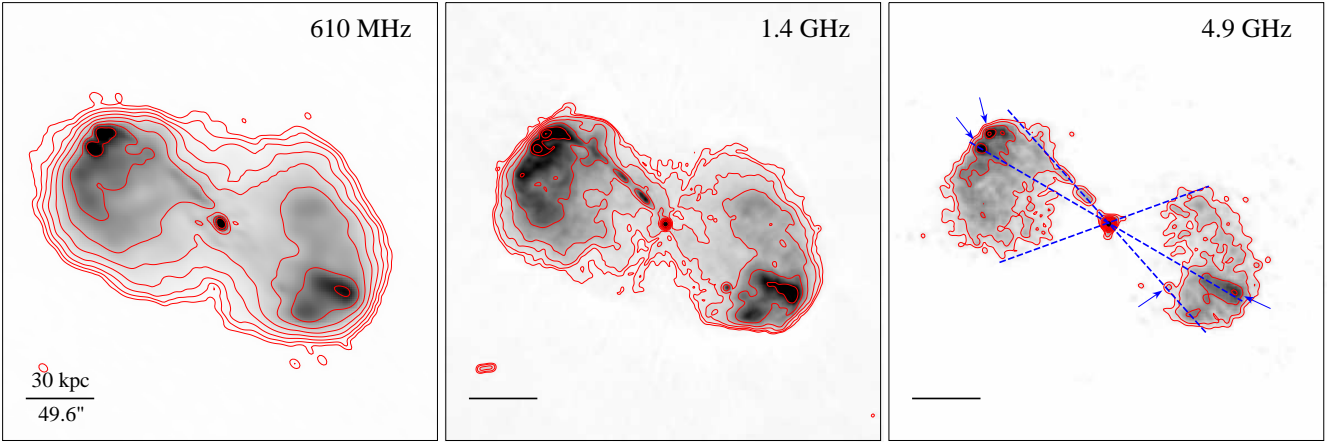}}
\vspace{-0.1cm}
\caption{\gmrt\ 610 MHz, \vla\ 1.4 GHz and 4.9 GHz images with an rms level of 100, 45, and 112 $\mu$Jy/beam (1$\sigma$) and contours of 3C~88 (see Section 2.2 for detail). The \gmrt\ contours start from 5 $\sigma$ and are spaced by a factor of 2. The 1.4 GHz contours start from 6 $\sigma$ and are spaced by a factor of 2. The 4.9 GHz contours start from 4.5 $\sigma$ to 625 $\sigma$ and the square-root spacing is adopted. The radio images at low and high frequencies show a similar morphology with extended lobes, a narrow jet, and four ``hotspot''-like features (marked by the blue arrows). The prominent northeastern jet bends near the end, towards the brightest ``hotspot'', while the other ``hotspot'' at the northeast is aligned with the ``hotspot'' at the southwest. The angle between these two blue lines is $\sim19\arcdeg$. We also draw a line along the edges of two radio lobes and the angle from it to the radio jet is $\sim68\arcdeg$. The radio morphology strongly suggests a jet reorientation projected from roughly east-west to northeast-southwest. The scale-bars in the middle and right panels are the same as the one in the left panel.	
	}
\label{radio}
\end{figure*}

\begin{figure*}
\includegraphics[scale=0.76]{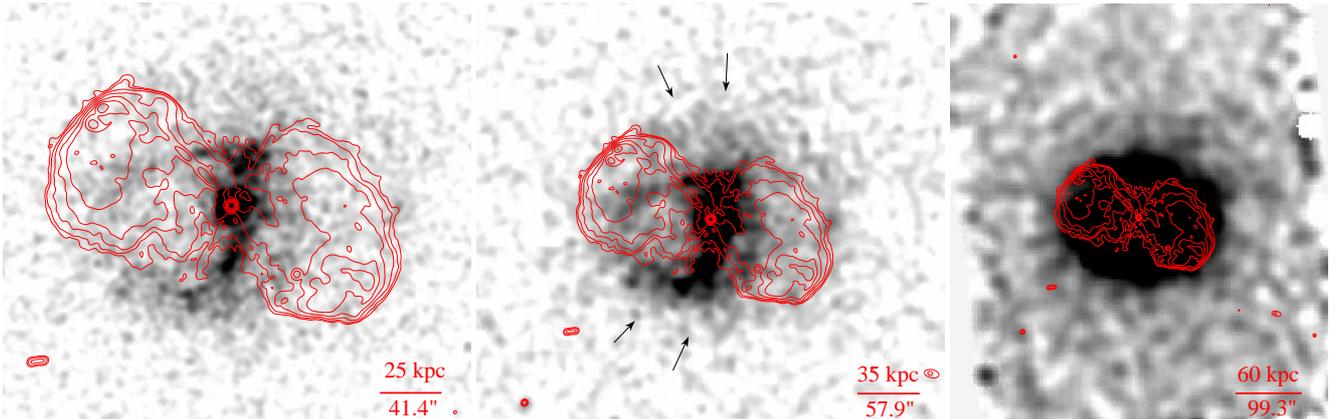}
\vspace{-0.4cm}
\caption{{\em Left}: The combined background-subtracted, exposure-corrected \chandra\ image of 3C~88 in the 0.5-3.0 keV band, smoothed with a two dimensional Gaussian 9-pixel kernel (0.492$''$/pixel) and overlaid with the \vla\ 1.4 GHz radio contours (in red as in Fig. \ref{radio}). The X-ray point sources have been removed and filled with the local background. The radio contours show a narrow jet, two bright radio lobes, and hotspots on each side of the radio nucleus. 
The prominent eastern X-ray cavity does not fully overlap with the brightest part of the radio lobe. 
	{\em Middle}: The same \chandra\ image as in the left panel smoothed with a two dimensional Gaussian 8-pixel kernel (0.984$''$/pixel). The image shows the prominent cavity to the east of the central AGN, the surrounding bright rims, and the edges associated with the cavity. X-ray decrement to the west is also likely in the position of the radio lobe.
	{\em Right}: The same \chandra\ image further zoomed out, smoothed with a two dimensional Gaussian 5-pixel kernel (3.936$''$/pixel), to show the elliptical edge just outside of the radio lobes. It is identified as a weak shock in this paper.
}
\label{image}
\end{figure*}
We reprojected and summed the count images, background images, and exposure maps respectively from three observations using the CIAO tool {\tt reproject\_image}. Fig. \ref{image} (left) shows the combined background-subtracted, exposure-corrected image overlaid with the \vla\ 4.9 GHz radio contours. 
Fig. \ref{image} (middle) is the zoomed-out image to show the prominent cavity to the east of the central AGN and the surrounding bright rims. Fig. \ref{image} (right) shows the further zoomed-out image with high contrast to highlight the shock edges.

The images show one prominent cavity at $\sim28$ kpc (projected distance) to the east of the nucleus with surrounding bright rims. 
The cavity does not fully overlap with the brightest part of the radio lobe. The current radio jets and hotspots are not aligned with
the cavity and have passed through it to larger radii, at least in projection. Beyond the bright rim of the cavity, there is a sharp 
surface brightness edge at $\sim65$ kpc to the south and north and at $\sim80$ kpc to the east and west. As shown in the next section, 
the gas temperature drops outward across the edge, which rules out the possibility that the edge is due to a cold front.

\begin{figure*}
\hbox{\hspace{5px}
\begin{tabular}{ll}
\includegraphics[scale=0.41]{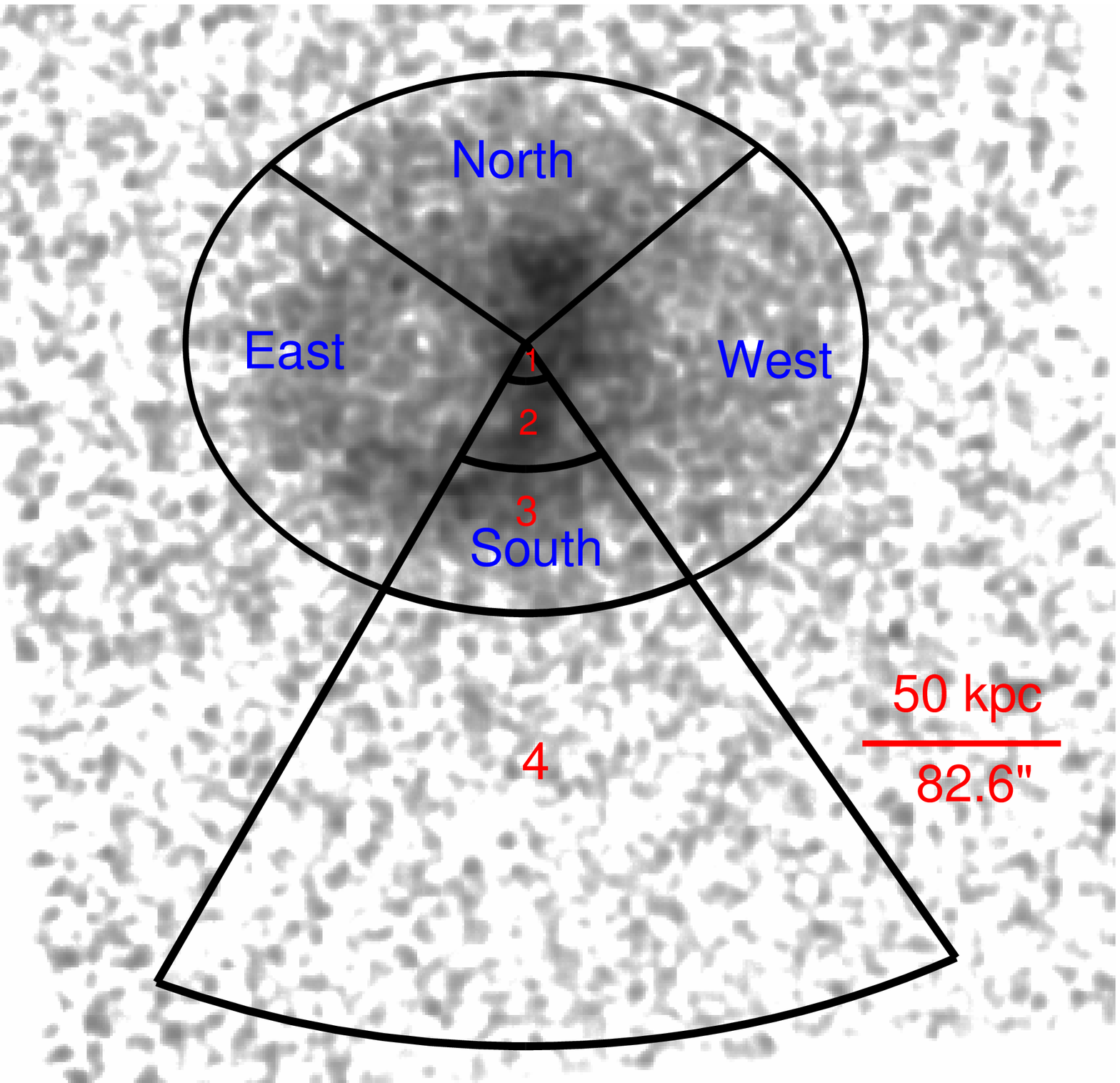}
&
\includegraphics[width=0.49\textwidth]{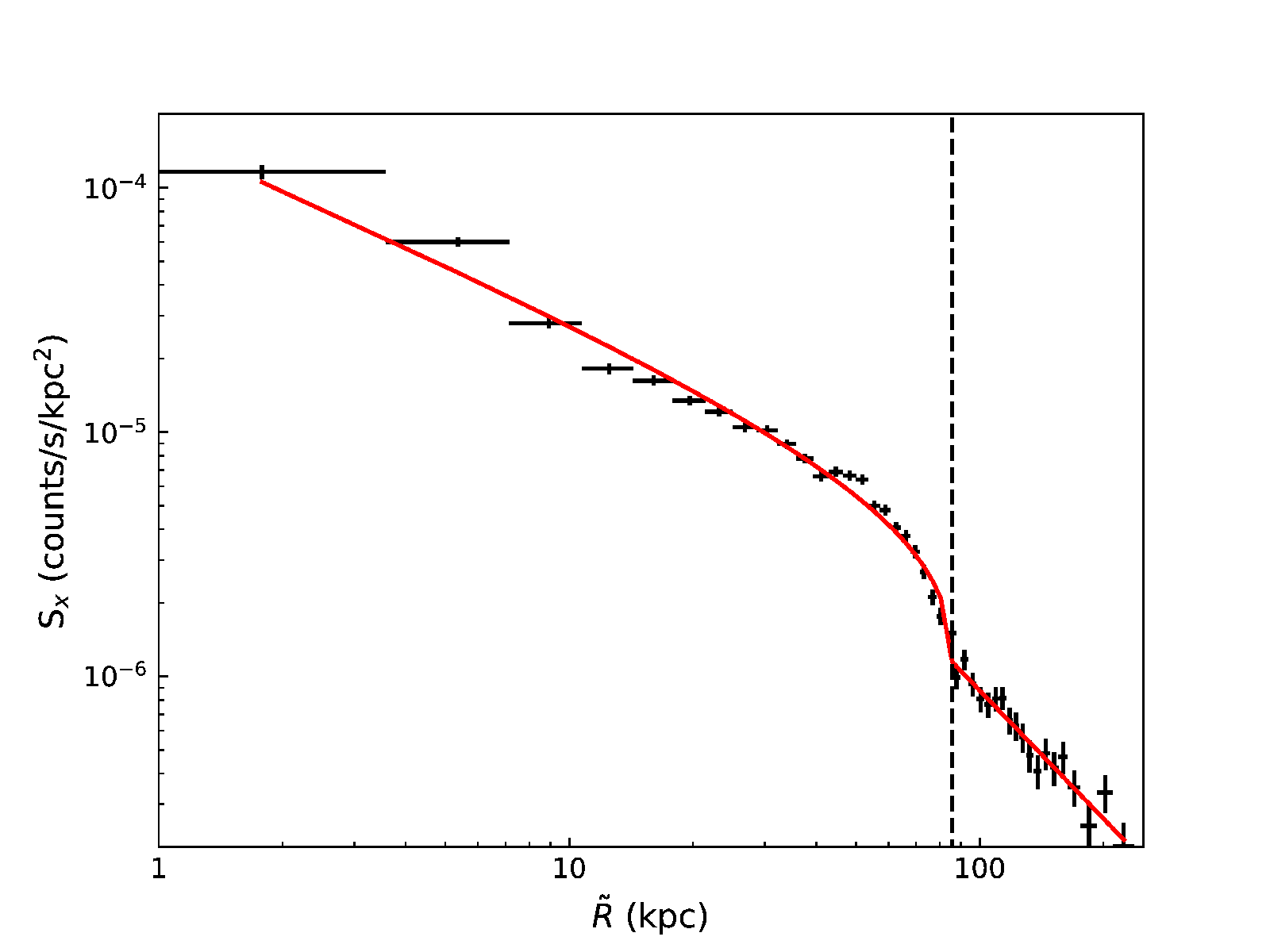}
\end{tabular}
}
\vspace{-5px}
\hbox{
\begin{tabular}{ll}
\includegraphics[width=0.49\textwidth]{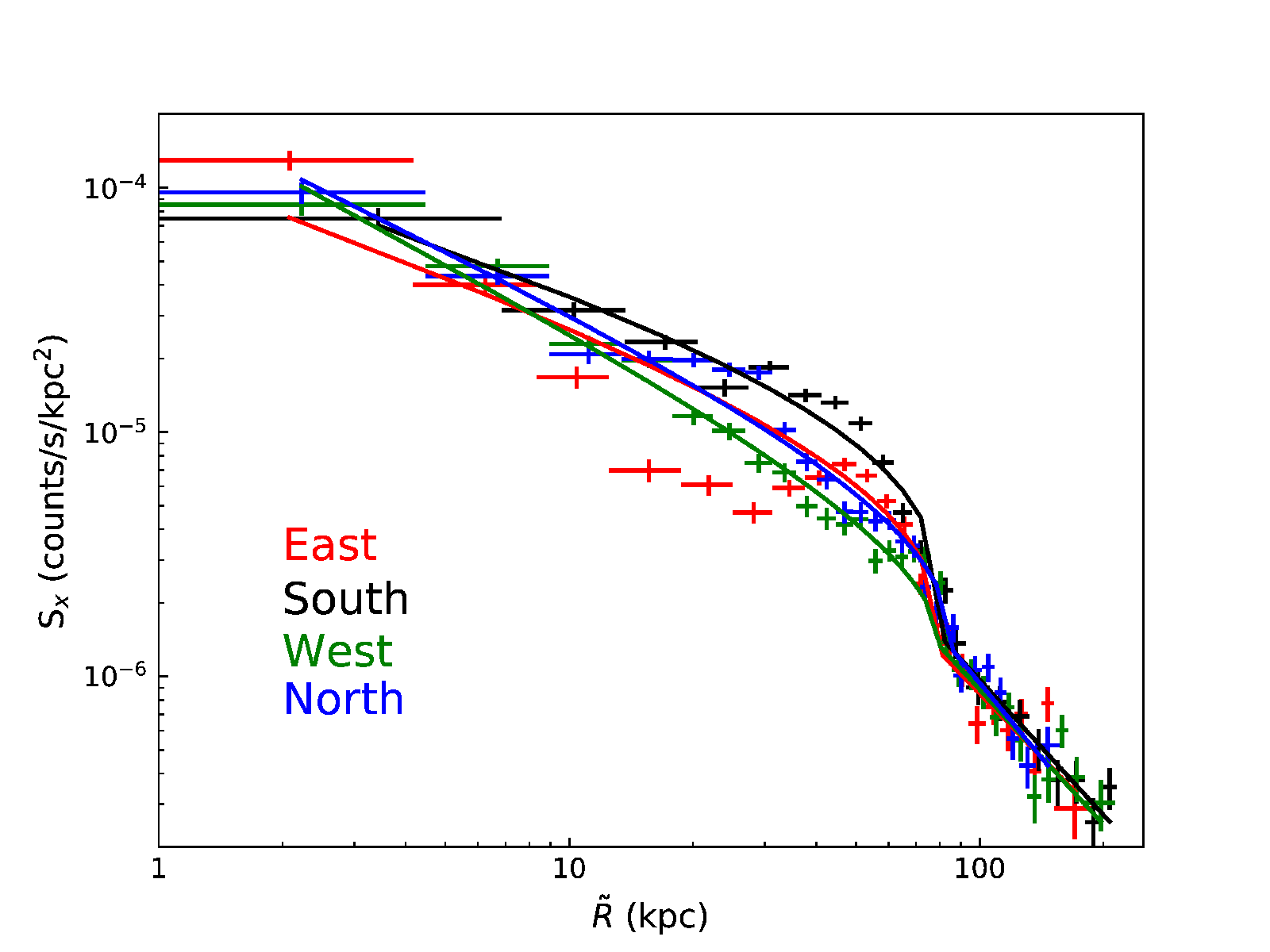}
&
\includegraphics[width=0.49\textwidth]{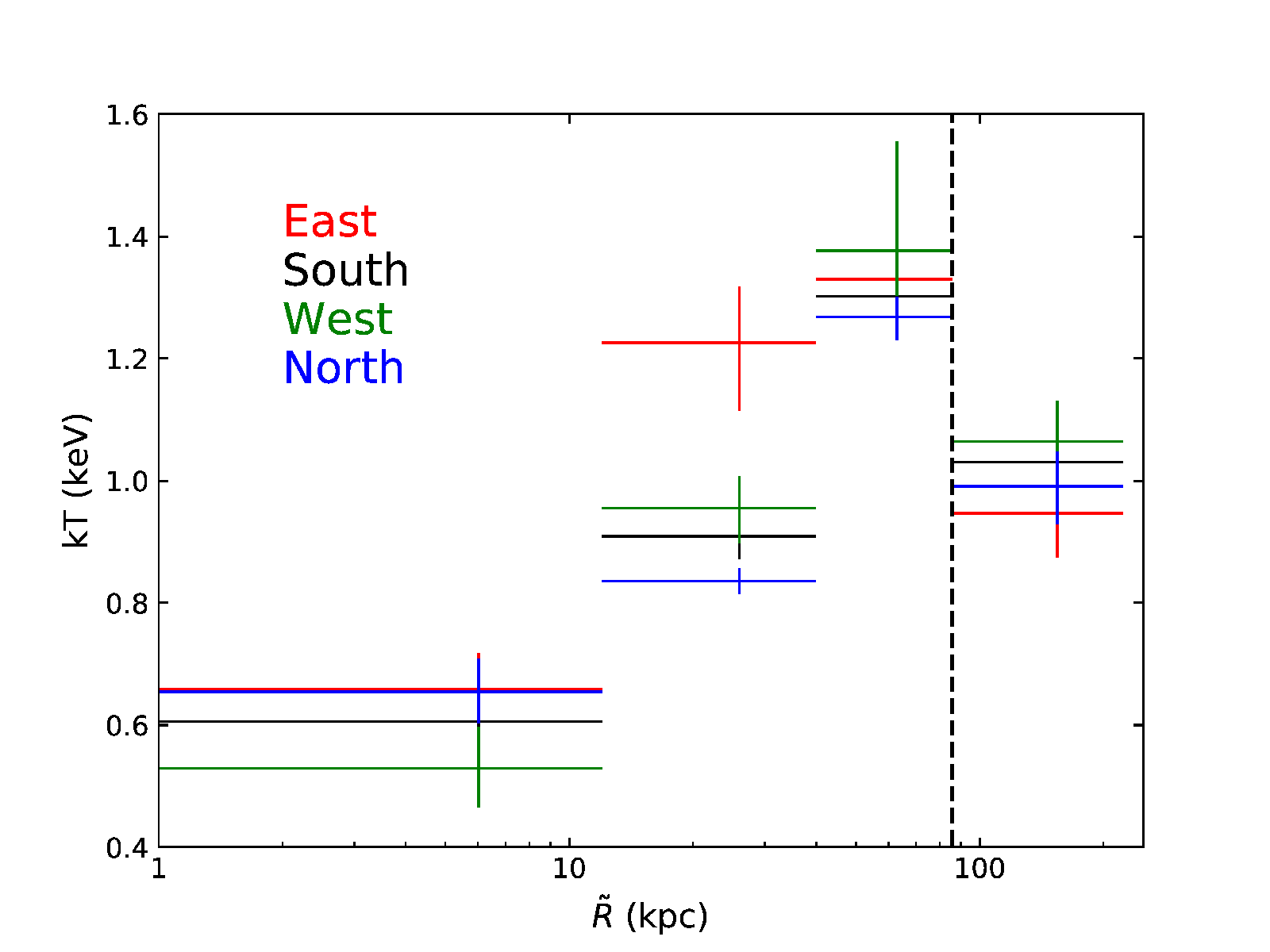}
\end{tabular}
}
\caption{Top left: The combined background-subtracted, exposure-corrected \chandra\ images of 3C~88 in the
	0.7-2.0 keV band, smoothed with a two dimensional Gaussian 10-pixel kernel (0.984\arcsec\/pixel). The surface brightness profiles
	are extracted in the same ellipse region from four wedges, with angles (measured counterclockwise from the west) of 
	145\arcdeg\ - 240\arcdeg\ for the eastern wedge, 240\arcdeg\ - 305\arcdeg\ for the southern wedge, 
	305\arcdeg\ - 400\arcdeg\ for the western wedge, and 40\arcdeg\ - 145\arcdeg\ for the northern wedge. 
	The ellipse shown in the image has a semi-major axis of 85.7 kpc and a major to minor axis ratio of 1.26.
	The regions marked in red are used for spectral analysis (we only show regions in the southern wedge). $\tilde{R}$ denotes 
	the semi-major axis of the ellipse.
	Top right: The averaged surface brightness profile across the edge in the 0.7-2.0 keV band.
	The red solid line shows the best-fit surface brightness model for an ellipsoidal emissivity edge. 
	The vertical dashed line shows the position (semi-major axis of $\sim85$ kpc) of the shock edge. 
	Bottom left: The surface brightness profiles in the 0.7-2.0 keV band for the four sectors shown in the top left panel. The solid lines show the best-fit surface brightness model. 
	Bottom right: the deprojected temperatures in region 1, 2, 3 and 4 for four wedges.
}
\label{SB}
\end{figure*}

To examine if the surface brightness edge surrounding the cavity is consistent with a shock discontinuity, we extracted the surface 
brightness profiles in the 0.7-2.0 keV band in the same ellipse region from four different wedges to match the shock edge.
The center is chosen to coincide with 3C~88's radio nucleus.
As shown in the top left panel in Fig. \ref{SB}, the angles (measured counterclockwise from the west) are 145\arcdeg\ - 240\arcdeg\ for the eastern wedge,
240\arcdeg\ - 305\arcdeg\ for the southern wedge, 305\arcdeg\ - 400\arcdeg\ for the western wedge, and 40\arcdeg\ - 145\arcdeg\ for the
northern wedge. We also extracted the averaged surface brightness profile within the full ellipse. We used the semi-major
axis of the ellipse as the X-axis for all the surface brightness profiles.
The particle background was subtracted using the normalized stowed background. 
We modeled the sky background (see section \ref{DataAnalysis}) and estimated a flux of $1.8\times10^{-7}$  count\,s$^{-1}$\,arcsec$^{-2}$
which is subtracted from the surface brightness. The radial bins were chosen to ensure a minimum of 200 source counts per
bin for the averaged surface brightness profile, and a minimum of 100 source counts per bin for those from
four wedges. From the surface brightness profile, the X-ray emission from the galaxy group is 
traced to $\sim220$ kpc. We can clearly see the depression of surface brightness in the eastern wedge from $\sim5-50$ kpc because of the prominent X-ray cavity
(Fig. \ref{SB} bottom left).

To fit the surface brightness discontinuity, we assumed a simple, self-similar ellipsoidal model for the X-ray emissivity,  
which is a power-law function of the elliptical radius (e.g., $\epsilon\propto r^{-p}$)
inside and outside of the discontinuity \citep{Sarazin16}. The best-fit surface brightness model is shown as the solid line in Fig. \ref{SB} top right panel for the averaged profile, and in the bottom left panel for 
the profiles from different wedges. For the eastern wedge, we ignored the data corresponding to the cavity. 
The best-fitting parameters are summarized in Table \ref{sbpfit} (notice that the larger errors in eastern and northern wedges are due to
the fewer surface brightness data points used in fitting). The edges are roughly 
located at $\sim80-86$ kpc. For the averaged surface brightness profile, the best-fit model gives the inner and outer 
power-law slope ($p_{\rm i}$ and $p_{\rm o}$ respectively)
of 1.37 and 0.87, and an emissivity jump ($R_{\rm em}=\epsilon_{i}/\epsilon_{o}$) of $3.11\pm0.30$ at the edge. Based on the best-fit model of the emissivity
distribution, we can obtain the related density distribution $n_{\rm e}(r) = [\epsilon(r)/\Lambda(T, Z)]^{1/2}$,
where $\Lambda(T, Z)$ is the X-ray emissivity function which depends on gas temperature $T$ and abundance $Z$. 
The density jump across the shock edge is $\rho_{\textrm{post}}/\rho_{\textrm{pre}} = 1.57\pm0.31$, where suffix pre and post denote 
quantities upstream and downstream of the shock. We calculated the Mach number of the shock based on the density jump
\citep[e.g.,][]{Landau59}:
\begin{equation}
	M^{2} = \frac{2C}{(\gamma+1)-(\gamma-1)C},
\end{equation}
where $M$ is the Mach number, $C=\rho_{\textrm{post}}/\rho_{\textrm{pre}}$ is the compression ratio and $\gamma=5/3$ is the adiabatic index. We obtained a Mach number
of $M = 1.39\pm0.23$.
\begin{table*}
\protect\caption{Results of Shock Edge Analysis}
\begin{tabular}{|c|c|c|c|c|c|c|c|c}
\hline 
region & $r_{\textrm{edge}}$   & $p_{\rm i}^{a}$    & $p_{\rm o}^{b}$ & $R_{\rm em}^{c}$    & Density Jump &  $M^{d}$ & Temperature Jump & $M^{e}$ \\
       & (kpc)        &            &         &        &              &        \tabularnewline
\hline 
East   & 81.26        & $0.79\pm0.32$ & $1.37\pm0.22$ & $3.72\pm1.38$ & $1.48\pm0.51$ & $1.33\pm0.36$ & $1.40\pm0.10$ & $1.41\pm0.10$ \tabularnewline
West   & 80.37        & $0.95\pm0.03$ & $1.42\pm0.13$ & $1.98\pm0.42$ & $1.79\pm0.19$ & $1.56\pm0.15$ & $1.29\pm0.14$ & $1.30\pm0.14$ \tabularnewline
South  & 82.45        & $0.75\pm0.06$ & $1.42\pm0.09$ & $4.80\pm0.85$ & $1.89\pm0.24$ & $1.64\pm0.20$ & $1.26\pm0.06$ & $1.27\pm0.06$ \tabularnewline
North  & 86.32        & $0.90\pm0.04$ & $1.52\pm0.25$ & $4.80\pm0.85$ & $1.87\pm0.47$ & $1.62\pm0.38$ & $1.28\pm0.08$ & $1.30\pm0.09$ \tabularnewline
Full   & 85.73        & $0.87\pm0.02$ & $1.37\pm0.06$ & $3.11\pm0.30$ & $1.57\pm0.31$ & $1.39\pm0.23$\tabularnewline
\hline 
\end{tabular}
\begin{tablenotes}
 \item
$a$: the inner slope of emissivity power-law; 
$b$: the outer slope of emissivity power-law; 
$c$: the emissivity jump at the edge; 
$d$: Mach number from density jump; 
$e$: Mach number from temperature jump; 
see section \ref{SBshock} for detail.\\
\end{tablenotes}
\label{sbpfit}
\end{table*}

\subsection{Spectral Analysis}
\label{spectraanalysis}
\subsubsection{Radial Spectral Analysis}
\label{specanalyse1}
\begin{figure*}
\begin{center}
\hbox{\hspace{5px}
\includegraphics[scale=1.0]{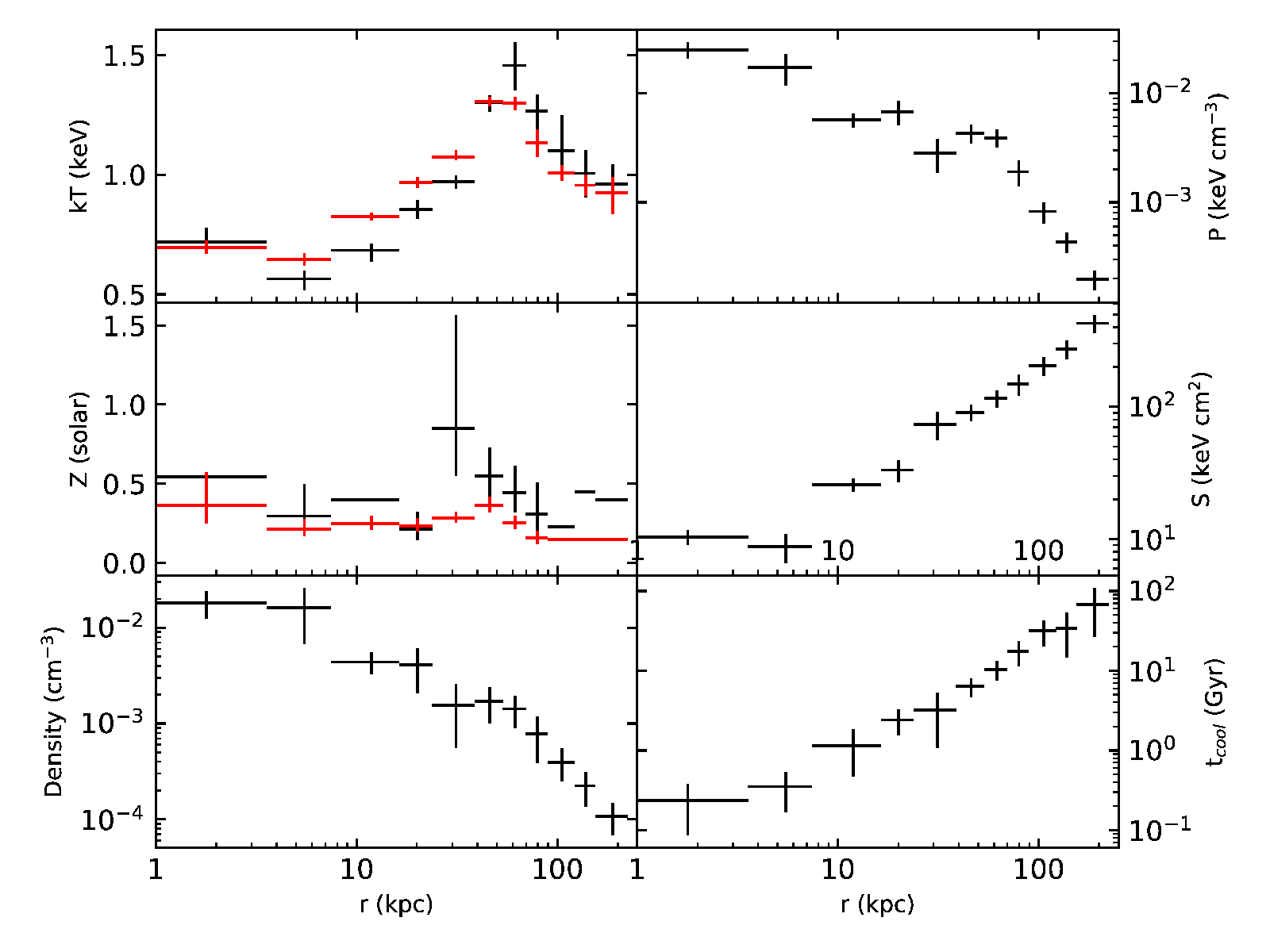}
}
\caption{\label{Tprofile}The azimuthally-averaged radial profiles of the deprojected temperature, abundance $Z$, electron density, pressure, entropy, and cooling time. For comparison, we also show the projected temperature and abundance profiles in red.}
\end{center}
\end{figure*}

\begin{figure}
\begin{center}
\hbox{\hspace{5px}
\includegraphics[scale=0.5]{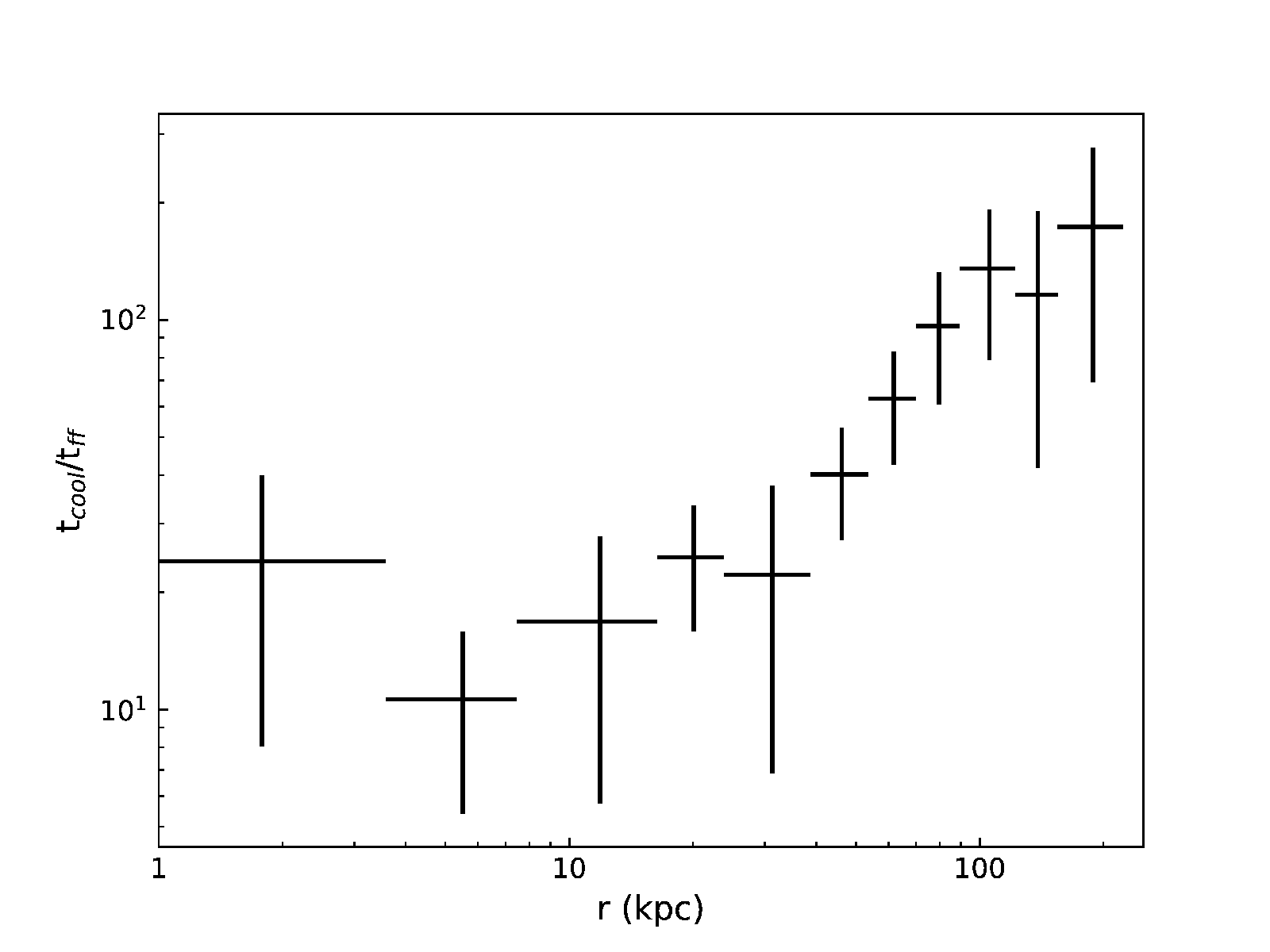}
}
\caption{\label{tratio}Radial profile of $t_{\textrm{cool}}$/$t_{\textrm{ff}}$.}
\end{center}
\end{figure}

\begin{figure*}
\begin{tabular}{ll}
\includegraphics[scale=0.33]{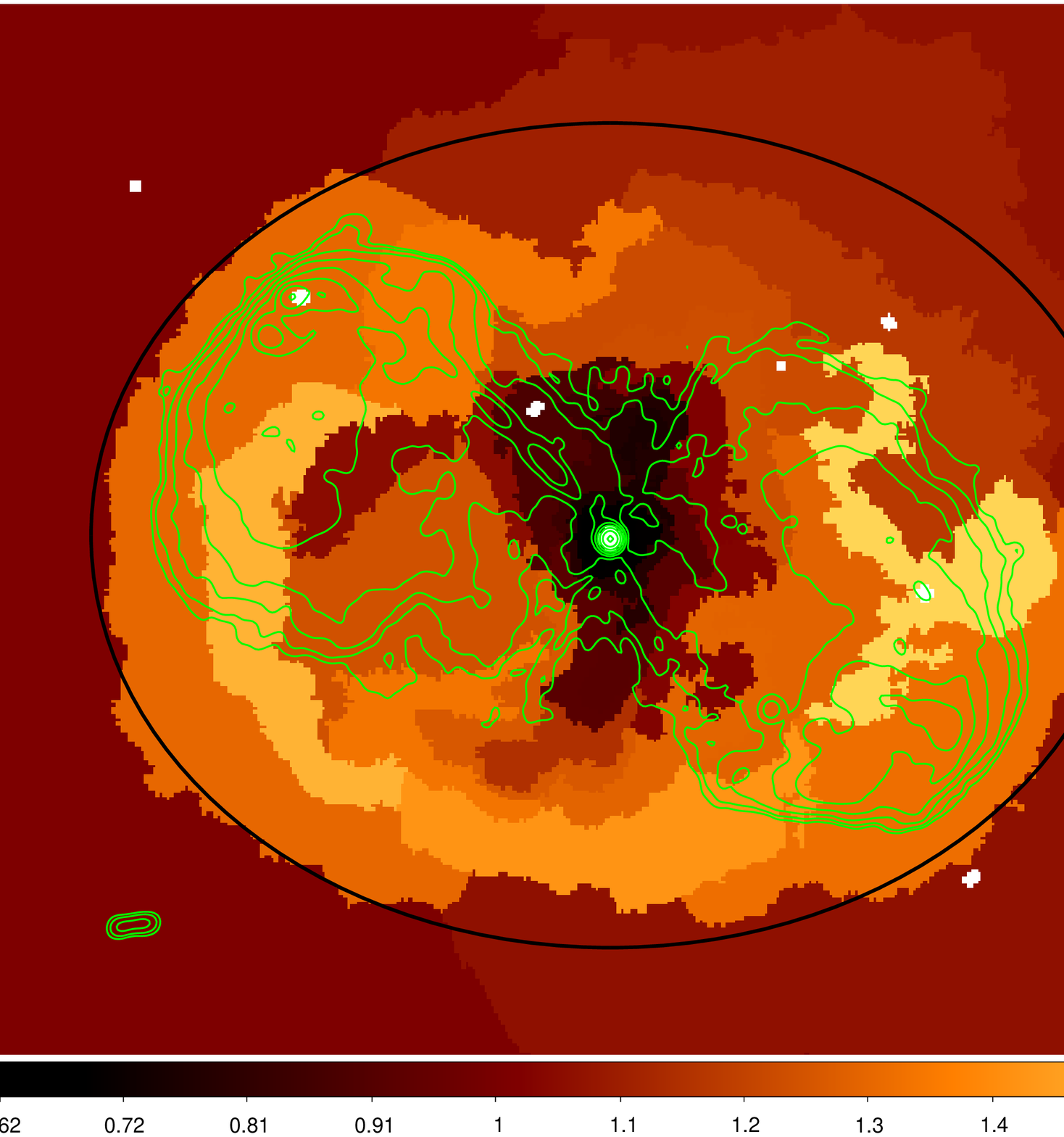}
&
\includegraphics[scale=0.33]{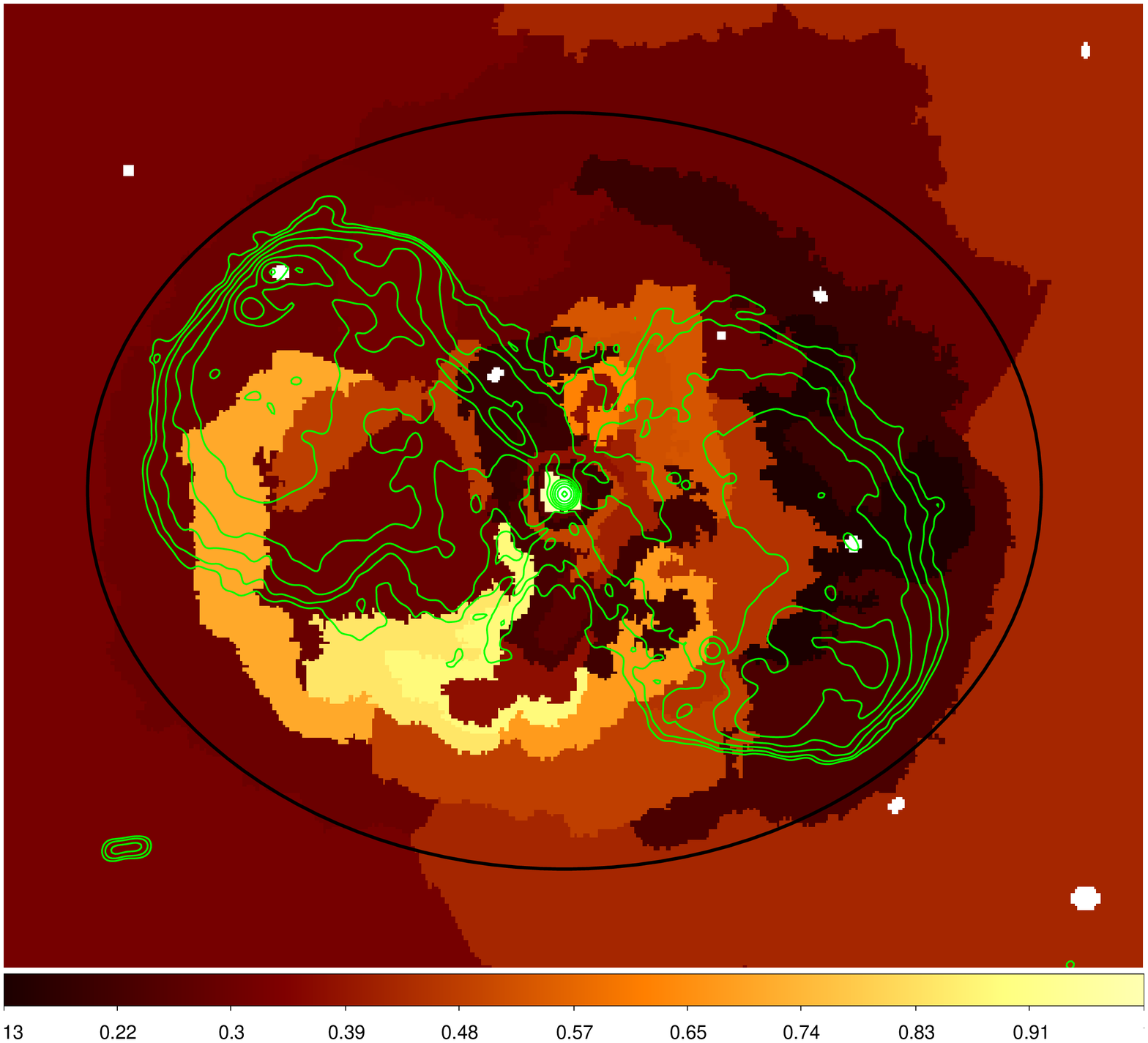}
\end{tabular}
	\caption{Temperature (left) and abundance (right) map of the central region overlaid with the \vla\ 1.4 GHz radio contours (in green) as in Fig. \ref{radio}. The black ellipse in the map is the same as in Fig. \ref{SB} to approximate the shock front.}
\label{TZ_map}
\end{figure*}

Spectra were extracted from the same circular annuli for the three observations separately using the CIAO tool {\tt specextract}. 
For each spectrum, the weighted 
response files and matrices were made, and a corresponding background spectrum was extracted from the same region from the 
normalized stowed background data. The sky background model obtained from the local background estimation (section \ref{DataAnalysis})
was scaled by the area of each region and subtracted.
The spectra were fitted in the energy range 0.5-5.0 keV with XSPEC 12.9 using the C-statistic.
The spectra from the same region from three observations were fitted simultaneously to an absorbed thermal APEC model, i.e., {\tt TBABS*APEC}.
The temperature, metallicity, and the normalization were allowed to vary freely. For regions where the abundances
cannot be well constrained due to the fewer counts, we fixed it at the best-fitting value or at the value obtained in nearby regions.

We focused our analysis on the ACIS-S3 chip and extracted a set of 11 circular annuli from the center to a radius of $\sim220$ kpc. The deprojected temperature profiles are 
shown in Fig. \ref{Tprofile} (top left). For temperature
deprojection, we adopted a model-independent approach \citep[e.g.,][]{Dasadia16} assuming spherical symmetry and the cluster 
temperature is uniform in each spherical shell. The approach is very similar to the ``onion-peeling'' method, but with 
a different way to estimate the emission contribution to each annulus from the outer spherical shells. We used
Monte Carlo simulations to populate the 3D shells and create the projected images where each pixel represents the number of events that fell
within each shell. By applying the respective masks (e.g., the included and excluded regions) to the simulated images, we can 
calculate the emission contribution factor from the outer spherical shell based on the sum of the pixel values. 
Using this method, we can account for point sources and chip gaps. We noted that our results are consistent with the results obtained by using the {\tt PROJCT} model in XSPEC.

The deprojected temperature profile increases from $\sim0.6$ keV in the center (within 10 kpc) up to $1.45$ keV at 
$\sim62$ kpc ($\sim0.1r_{180}$, $r_{180}$ is the virial radius and $r_{\Delta}$ is defined as the radius where the 
mean mass density equals to $\Delta$ times of the critical density.), 
then declines towards large radii. This behavior is very similar to the temperature profiles observed in other galaxy groups 
\citep[e.g.,][]{Gastaldello07,Sun09}. 
We noticed that the temperature peak is located just inside the shock region, which may complicate the detection of the
actual temperature jump due to the shock itself, as found in other systems, e.g., in Hydra A \citep{Gitti11}.
The abundances (Fig. \ref{Tprofile} middle left) are not well constrained in a few annuli and therefore are fixed.
The abundance within $\sim20$ kpc is relatively flat at a low value of $\sim0.3$. The low abundance at the center of cooling flow
clusters is mainly due to fitting multiple temperature spectra with a single thermal 
model \citep[the so-called ``Fe bias'' effect, ][]{Buote00a}. 
The abundance in the fifth data bin where the cavity is located shows a jump, but with a large uncertainty.

We also present various quantities in Fig. \ref{Tprofile}. 
The electron density was obtained directly from the deprojected 
normalization of the APEC component assuming $n_{\rm e}=1.22n_{\rm H}$, where $n_{\rm e}$ and $n_{\rm H}$ are the electron and proton densities. 
A dip is shown in the density profile at $\sim28$ kpc corresponding to the location of the cavity (Fig. \ref{Tprofile} bottom left).
Based on the deprojected gas temperature and density, we calculated the pressure in each annulus as $P=nkT$ (where $n=1.92n_{\rm e}$) and 
the entropy as $S=kT/n_{\rm e}^{-2/3}$.
As shown in Fig. \ref{Tprofile} middle right, the entropy profile declines all the way to the center.
The entropy profile can be well represented by a constant plus a power-law as $S=S0+S1(r/100 \textrm{ kpc})^{p}$,
with the best-fitting slope of $1.19\pm0.09$ and the central entropy floor $S0=7.31\pm1.68$ keV cm$^{2}$.
We found the best fitting entropy slope of $0.79\pm0.07$ between 30 kpc$ - 0.15r_{500}$, which is consistent with the values found in galaxy groups
and clusters \citep[e.g.,][]{Sun09}, and the slope of $1.19\pm0.05$ between $0.15r_{500} - r_{2500}$ (e.g., 65 - 206 kpc).

We calculated the gas cooling time as
\begin{equation}
	t_{\rm cool}= \frac{3P}{2n_{\rm e}n_{\rm H}\Lambda(T,Z)},
\end{equation}
where $\Lambda(T,Z)$
is the cooling function determined by the gas temperature and metallicity. The cooling time at the center of 3C~88 is below 0.5 Gyr 
(Fig. \ref{Tprofile} bottom right). Conventionally, the cooling radius is defined as the radius where the cooling time is shorter
than 1 Gyr or the lookback time to $z=1$ (7.7 Gyr), which corresponds to a radius of $\sim$ 11 kpc or 50.5 kpc for 3C~88. We estimated a 
bolometric X-ray luminosity of $3.3\times10^{41}$ erg s$^{-1}$ within the cooling radius (11 kpc, corresponding to the cooling time of 1 Gyr) 
for this strong cool core group.

Assuming hydrostatic equilibrium and spherical symmetry, we calculate the gravitational acceleration, $g$, as
\begin{equation}
	g=\frac{d\Phi}{dr} = -\frac{1}{\rho}\frac{dP}{dr},
\end{equation}
where $\Phi$ is the gravitational potential, $\rho=1.92n_{\rm e}\mu m_{\rm p}$ is the particle density, $\mu=0.60$ is the mean particle weight and
	$m_{\rm p}$ is the proton mass. To obtain a smooth pressure profile, we fitted it with a generalized NFW
model, where $P\propto r^{-3}$ at large radii and the inner slope is allowed to vary. We found the best-fitting inner slope of $0.58\pm0.11$. 
Once the gravitational acceleration is determined, we calculated the free fall time 
$t_{\rm ff}=\sqrt{\frac{2r}{g}}$ \citep[e.g.,][]{Gaspari12a}. As shown in Fig. \ref{tratio}, the minimum value of
$t_{\rm cool}/t_{\rm ff}$ at the center of 3C~88 is close to the threshold ($\sim10-20$), below which the hot gas becomes thermally unstable 
\citep{Sharma12,Gaspari12a,Voit15a}.
We also checked the turbulent eddy time scale $t_{\rm eddy}$ (defined as $2\pi\frac{r^{2/3}L^{1/3}}{\sigma_{v,L}}$, where $L$ is
the injection scale, $\sigma_{v,L}$ is the velocity dispersion at the injection scale) since the locus of the eddy time and cooling time
provides a robust criterion for the physical state of the hot gas \citep[e.g.,][]{Gaspari18}. We found
a minimum $t_{\rm cool}/t_{\rm eddy} \approx 1.0$, consistent 
with the presence of multiphase gas (section \ref{warmgas}) 
with an origin related to the turbulent CCA condensation process \citep[e.g.,][]{Gaspari18}.

\subsubsection{Temperature and Abundance Map}
To study the thermal structure of the group, we performed a 2D spectroscopic analysis using contour binning \citep{Contourbinning}.
We generated 32 regions with a signal-to-noise ratio of 50 for 3C~88 using counts image in 0.7-2.0 keV band. We used the model {\tt TBABS*APEC} to model
the gas emission in each region in 0.5-5.0 keV band. The resulting projected temperature and abundance maps are shown in 
Fig. \ref{TZ_map}. 
The temperature map shows the high temperature region surrounding the eastern cavity, at the location of the shock edges. The
abundance map shows low abundance regions close to the central AGN, which again is possibly due to the ``Fe bias'', and 
the high abundance regions surrounding the cavity, which we will discuss later in section \ref{rimsarms}.

\subsubsection{Shock Spectral Analysis}
In section \ref{SBshock}, the shock features are derived mainly from the surface brightness profile. Here we present 
the spectral analysis of the shock.
As shown in Fig. \ref{SB} (top left), each wedge was divided into four small regions (1, 2, 3 and 4).
For each region, the spectra were extracted from individual observations and were fitted simultaneously in the 0.5-5.0 keV band.
The deprojected temperature values were shown in Fig. \ref{SB} (bottom right), obtained by assuming
ellipsoidal symmetry with one of the principal axes along the line-of-sight.
We estimated the Mach number of the shock based on the temperature jump \citep[e.g.,][]{Landau59,Sarazin16}:
\begin{equation}
	M^{2} = \frac{(8\frac{T_{2}}{T_{1}}-7)+[(8\frac{T_{2}}{T_{1}}-7)^{2} + 15]^{1/2}}{5},
\end{equation}
where $M$ is the Mach number, $T_{1}$ and $T_{2}$ are the pre- and post-shock temperature, and we assumed $\gamma=5/3$. 
As shown in Table \ref{sbpfit}
we obtain Mach numbers of $1.41\pm0.10$, $1.27\pm0.06$, $1.30\pm0.14$, and $1.30\pm0.09$ for eastern, southern, western, and northern
wedges, respectively. 

\section{The X-ray Cavity}
\label{cavity}
\begin{figure}
\includegraphics[scale=0.32,angle=-90]{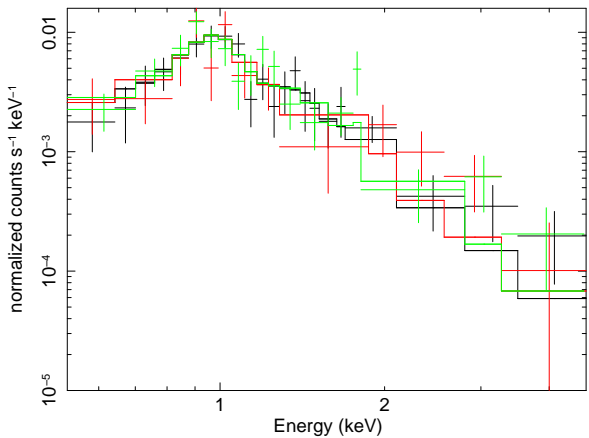}
\caption{Spectra at the position of the eastern cavity from three observations, with the best-fit single temperature model (see more detail in section \ref{cavity}).
}
\label{spec_cavity}
\end{figure}

In Fig. \ref{image} (left) the smoothed \chandra\ X-ray image of 3C~88 shows a pronounced cavity to the east of the core, which
is also evident from the surface brightness profile extracted from the eastern wedge as in Fig. \ref{SB}. The corresponding cavity to
the west is not significant. 

We estimated the lower limit of the total jet power based on the enthalpy and buoyant rise time of the cavities \citep{Churazov01}, to ensure the direct comparison with previous studies. We assume the cavity has a spherical geometry centered at 3:27:57.327,+2:33:33.623 with a radius of 0.633\arcmin. The projected distance of the center of the eastern cavity to the nucleus is $\sim28$ kpc, with a radius of 23 kpc. The buoyant rise time of the bubble is calculated at its terminal velocity $\sim(2gV/SC)^{1/2}$, where $V$ is the volume of the bubble, $S$ is the cross section of the bubble, and $C=0.75$ is the drag coefficient \citep{Churazov01}. The gravitational acceleration is calculated as $g\approx 2\sigma^{2}/R$, with the central stellar velocity dispersion of 189 km s$^{-1}$ \citep{Smith90}. Thus, the age of the bubble to its present position is $6.0\times10^{7}$ yr. The total cavity enthalpy ($H = 4PV$, where $P$ is the azimuthally-averaged pressure at the radius of the center of the cavity) is calculated from the deprojected temperature and the electron density profile. The total enthalpy and the cavity power for the eastern are $3.8\times10^{58}$ erg and $2.0\times10^{43}$ erg s$^{-1}$. We noted that the cavity power to the cooling luminosity ratio is $\sim61$, showing that the mechanical power is sufficient to offset the cooling. If we instead use $g$ at the center of the cavity derived from section \ref{spectraanalysis}, the age of the bubble is $\sim$ 40\% lower and the resulting cavity power is $\sim$ 60\% higher.

For comparison, we estimated the jet power using another technique based on the observed parameters of the jet terminal hotspots,
which is calculated as \citep[e.g.,][]{Godfrey13}
\begin{equation}
	Q_{\rm hs} = Ac\frac{B_{\rm eq}^{2}}{8\pi}\times g_{\rm hs},
\end{equation}
where $A$ is the area of the jet hotspots, $c$ is the speed of light, $B_{\rm eq}$ is the equipartition magnetic field strength \citep{Worrall09}, 
and $g_{\rm hs}$ is the normalization factor which can be empirically determined.
From the \vla\ 4.9 GHz radio map, we estimated the radii of the hotspots on both sides as $\sim2.4$ kpc. 
We estimated an equipartition magnetic field of $\sim5$ $\mu$G based on the radio data using equation 2 in \citet{Miley80}
(We need to be aware of the uncertainties in this determination). 
With a $g_{\rm hs}$ factor of 2 \citep[e.g.,][]{Godfrey13}, we estimated a jet power of $\sim2.1\times10^{43}$ erg s$^{-1}$.
As the plasma in the hotspots is likely to be relativistic, their properties are thought to be variable on timescales comparable
to the light crossing time of a hotspot. While the jet power estimated from above equation
could vary on a timescale much shorter than that used to estimate the cavity power, we note that jet power is about the same as the cavity power.

The spectra in Fig. \ref{spec_cavity} were extracted from each observation of the eastern cavity region (see Fig. \ref{image}),
together with response files and background spectra.
We fitted the spectra with a one temperature model as {\tt TBABS*APEC} and obtained the best fitting projected temperature 
of $1.22^{+0.06}_{-0.07}$ keV and the abundance of $0.22^{+0.09}_{-0.06}$ solar.
Although the one temperature model provides a good fit to the spectra, it is lower than the data above $\sim3$ keV as shown in Fig. \ref{spec_cavity}. Thus,
we added another thermal component to place constraints on any hot gas in
the cavity. We fixed the temperature of the first thermal component to 1.05 keV at 
the radius of the cavity obtained from the projected fitting to account for the
projected emission. The second component accounts for any thermal emission inside the cavity. The temperature
is constrained between 1.5 and 10 keV. The metallicities of the two components are tied together. 

We found that adding a hotter component improves the fit with a significance of 99\% using F-test, implying the existence of hot gas in the cavity. 
Assuming the hot gas in the cavity is in pressure balance with the cooler gas,
we can place constraints on the volume-filling factor based on the temperature and normalizations of the cool
and hot gas. We found that, as the temperature of the second component increases, the metallicity increases and
reaches above $\sim1.0$, which is a typical value for group centers. This may be the result of ``Fe bias'', which typically occurs when
fitting a single temperature model to the spectrum of the multiphase gas. The volume-filling factor increases rapidly from 0.72 for 
1.5 keV to 0.99 for temperatures greater than $\sim5.5$ keV.

\section{Warm gas in 3C~88}
\label{warmgas}
\begin{figure*}
\includegraphics[scale=0.52]{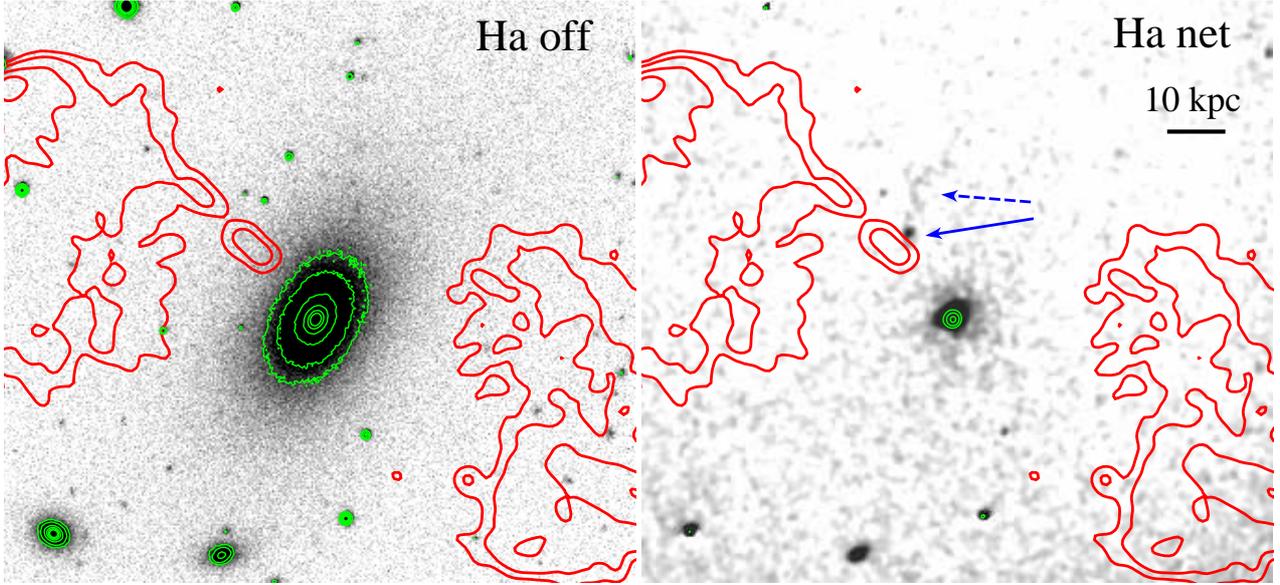}
\caption{Optical continuum (H$\alpha$ off) and net H$\alpha$+[NII] images of the center with the 3.5m {\em APO}
telescope. The \vla\ 4.9 GHz contours (red, same as the ones shown in Fig. \ref{image}) and optical contours (green) of the bright
	part are also shown. There is an H$\alpha$+[NII] extension beyond the central nucleus. Blue arrows in the right panel mark a compact 
	emission-line feature (solid blue line) and a potential emission-line diffuse feature (dashed blue line) close to the NE jet, more than 16 kpc away from the nucleus.
}
\label{Ha_source}
\end{figure*}

Narrow-band imaging data on 3C~88 were taken on December 16, 2009, with the Seaver Prototype Imaging camera
({\em SPIcam}) on the Astrophysical Research Consortium 3.5-meter telescope at the Apache Point Observatory.
SPIcam is a general purpose, optical imaging CCD camera with a field of view 4.78$'$ square.
Two 900 sec H$\alpha$+[NII] observations were taken with the NMSU-677.6 filter (central $\lambda$: 6776 \AA,
width: 75\AA). Two 540 sec observations with the NMSU-665 filter (central $\lambda$: 6650 \AA,
width: 80\AA) were taken for the continuum (or H$\alpha$ off). Both filters are 2 inch square, so only the central $\sim 3'$
square is unvignetted. The seeing was $\sim 1.1''$ and the sky was photometric during our observations.

We follow the standard CCD analysis for the {\em SPIcam} data. Dome flats were used and the spectrophotometric standard was Feige34. The Galactic extinction was 0.259 mag for the NMSU-665 filter and 0.252 mag for the NMSU-677.6 filter, with the extinction law from \citet{Fitzpatrick99} and assuming $R_{\rm V}$ = 3.1.
A software issue with the {\em SPIcam} shutter in 2007 - 2010 adds 0.7505 sec to the requested exposure.
This was identified at the end of 2010 and was corrected for our analysis.

For continuum subtraction, we generally follow the isophote fitting method described in \citet{Goudfrooij94}, by studying the H$\alpha$ / off ratio profile. The net H$\alpha$+[NII] emission of the galaxy is shown in Fig. \ref{Ha_source}. 3C~88 hosts a nuclear source with H$\alpha$+[NII]. A diffuse extension, beyond the nuclear source, is detected to at least 5 kpc from the nucleus.
The presence of multiphase gas with $\sim$5 kpc extent (and fairly low $t_{cool}$ compared with the free-fall/eddy timescales; Sec. \ref{specanalyse1}) can 
be understood in the CCA scenario \citep[cf.,][]{Gaspari18}, in which cold/warm gas originates from the in-situ condensation cascade and recurrently triggers the AGN jets.
Further away from the nucleus, a compact knot is detected 16 kpc NE of the nucleus, with potentially a faint filament to its north.
Within the central 6 kpc radius, where most of the central diffuse emission lies, we measure
a total H$\alpha$ luminosity of 6.6$\times$10$^{39}$ erg s$^{-1}$, assuming no intrinsic extinction.
The small blob, 16 kpc NE of the nucleus, has a total H$\alpha$ luminosity of 5.3$\times$10$^{39}$ erg s$^{-1}$,
if it is associated with 3C~88.
The following line ratios are assumed, [NII]6583/H$\alpha$ = 1.5 and [NII]6548/[NII]6583=1/3, which are consistent with the typical values in \citet{Goudfrooij94}.
Deeper narrow-band imaging data with good seeing and spectroscopic data are required to better understand the state of warm gas in this system. Nevertheless, 3C~88 does host some warm gas near the nucleus and the radio jet.

Since 3C~88 was not observed by {\em Herschel} or {\em Spitzer}/{\em MIPS} and it was not detected in
{\em Galex} FUV AIS data, we estimate its SFR solely from the {\em Wise} 22$\mu$m data.
Based on the SFR calibration by \citet{Lee13},
the SFR for 3C~88 is 0.96 $M_{\odot}$\,yr$^{-1}$ for the Kroupa IMF. However, one can only view this estimate as an upper limit,
as the nuclear emission may dominate the 22$\mu$m emission. Star formation in 3C~88 is at most weak.

\section{Discussion}
\label{discussion}
\subsection{The Central AGN}
\begin{table*}
\protect\caption{Spectral fits of the central AGN}
\begin{tabular}{|l|c|c|c|c|c|l|}
\hline
Model               & $\Gamma_{1}$$^a$            & $kT/\Gamma_{2}$$^b$        & $N_{\rm H, intr}$$^c$        &  $L_{0.5-2.0}$$^d$         & $L_{2-10}$$^e$           & $[C, C_{e}, C_{\sigma}]$$^f$ \\
&                         & (keV)                  & ($10^{22}$ cm$^{-2}$)  &  ($10^{40}$erg s$^{-1}$)      & ($10^{41}$erg s$^{-1}$)       &   \tabularnewline
\hline
tbabs*POW             & $0.60^{+0.04}_{-0.04}$  &                        &                        &                        & $5.71^{+0.21}_{-0.21}$ & [157.8, 131.0, 16.3] \tabularnewline
tbabs*(ztbabs*POW+APEC) & $1.40^{+0.14}_{-0.14}$  & $0.73^{+0.07}_{-0.14}$ & $1.6^{+0.3}_{-0.3}$ & $1.78^{+0.23}_{-0.22}$ & $4.57^{+0.20}_{-0.19}$ & [115.9, 131.1, 16.0]  \tabularnewline
tbabs*(ztbabs*POW+APEC) & $1.70$                  & $0.77^{+0.06}_{-0.07}$ & $2.1^{+0.2}_{-0.2}$ & $1.97^{+0.23}_{-0.21}$ & $4.19^{+0.15}_{-0.14}$ & [120.3, 131.4, 15.9]  \tabularnewline
tbabs*(ztbabs*POW+POW)  & $1.70$                  & $1.13^{+0.22}_{-0.19}$ & $3.1^{+0.4}_{-0.4}$ &                        & $2.79^{+0.32}_{-0.36}$ & [118.3, 131.6, 16.4]  \tabularnewline
\hline
\end{tabular}
\begin{tablenotes}
 \item
	 $a$: the photon index for the first power-law component;
	 $b$: the temperature for the thermal component (we assumed an abundance of 0.3 $Z_{\odot}$. The fits are not sensitive to the assumed abundance value.) or the photon index for the second power-law component; 
	 $c$: the column density for the intrinsic absorption;
	 $d$: the rest-frame 0.5-2.0 keV luminosity for the thermal component;
	 $e$: the rest-frame 2-10 keV luminosity for the power-law component;
	 $f$: $C, C_{e}, C_{\sigma}$ are the fitted C-statistic, its expected value, and its standard deviation computed based on \citet{Kaastra17}. 68\% of the time the acceptable spectral
	 models should have $-1< \frac{C-C_{e}}{C_{\sigma}} < 1$.
\end{tablenotes}
\label{tab_pointsource}
\end{table*}

\begin{figure}
\includegraphics[scale=0.5]{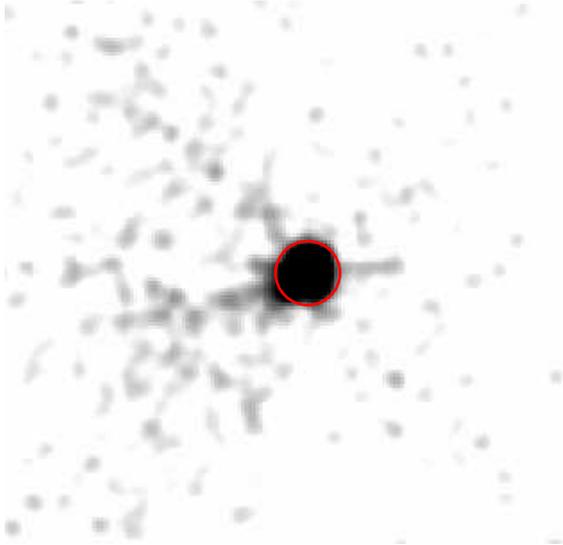}
\caption{The 0.7-7.0 keV subpixel (0.0984\arcsec/pixel) image of the central region, smoothed with a two dimensional Gaussian 4-pixel kernel. The spectrum of the central AGN is extracted from a circle with a radius of 2\arcsec\ that is shown in red. Despite the limited statistics of the data, some features are revealed around the center, including a linear feature to the northeast on the same direction of the radio jet (inner X-ray jet?) and extension to the east and west.
}
\label{subpixel}
\end{figure}
Fig. \ref{subpixel} shows the subpixel (0.0984\arcsec/pixel) \chandra\ X-ray image of the central region. We extracted spectra of the central X-ray source from each observation in a circular region with a radius of 2\arcsec, 
and the background spectra from an annulus from 2\arcsec\ to 4\arcsec. The spectra from the three observations were fitted simultaneously between 0.4 and 8.5 keV.
The nuclear spectrum is flat and the power-law photon index is $\sim$ 0.6 from a simple power-law model (Table \ref{tab_pointsource}).
We confirmed that pileup is negligible even for the brightest ACIS pixel.
As the local background is simply scaled by the solid angle, the residual thermal emission is expected in the nuclear spectrum. As the spectrum is flat, simply adding a thermal component results in a negligible contribution. We attempted models with an obscured AGN plus a soft component, which indeed improves the fits (Table \ref{tab_pointsource}).
If the canonical value of 1.7 for the power-law index of the AGN is preferred, 3C~88's AGN is obscured with a moderate intrinsic absorption column density of $\sim 2\times10^{22}$ cm$^{-2}$.  The rest-frame 2-10 keV luminosity of the AGN is $\sim 4\times10^{41}$ erg s$^{-1}$.

We examined the variability of the central point source based on the flux from the 0.5-10 keV band. As shown in Table \ref{tab_obs} the time separations for three observations 
are $\sim7$ and $\sim1$ day. Since the AGN variabilities have different timescales, we also include the old, short
ACIS-I observation (Obs. ID 9391) taken more than one year earlier. We found that there are small variations between the four 
observations and the 0.5-10 keV flux changes by $\sim20$\%.
We also checked the possible X-ray emission associated with the radio features, e.g., radio jet and the hot spots. Like \citet{Massaro15}, we 
chose a circular region for the hot spots and a box region for the radio jet to measure the observed X-ray emission. Background regions of the same shape and size, 
were chosen to avoid emission from other sources, with one box for the radio jet, and two circles for the hot spots. 
X-ray emission from the radio jet ($\sim25\arcsec$ northeast to the nucleus) is detected with a significance of $\sim4.0\sigma$, and 
with a significance level of $\sim4.7\sigma$ from the eastern hot spot \citep[$\sim109\arcsec$ to the nucleus;
note that this feature is marked as ``jet knot'' in][]{Massaro15}.
There is no detection of X-ray emission associated
with the western hot spot. The hardness ratio (2.0-7.0 keV flux divided by 0.5-2.0 keV flux) of the eastern hot spot is $\sim0.17$, which corresponds to
a power-law index of $\sim2.5$ assuming an absorbed power-law model. We estimated a 0.5-10 keV X-ray luminosity of $6.0\pm1.3\times10^{39}$ erg s$^{-1}$ for the eastern
hot spot. Assuming the same spectral shape, we obtained an upper limit luminosity of $\sim10^{39}$ erg s$^{-1}$ for the western hot spot.

Based on the core flux density measured by the \vla\ \citep{Morganti93} at 5 GHz, we calculated a radio luminosity at 5 GHz of 
$L_{\textrm{R}}=1.68\times10^{40}$ erg s$^{-1}$. Using the black hole fundamental plane \citep{Gultekin09}, or the relation between the radio luminosity at 5 GHz, 
the X-ray 2-10 keV luminosity, and the black hole mass, we estimated a black hole mass of $\sim 7.4\times10^{8}$ M$_{\odot}$, which is comparable to the black hole mass of $\sim5.0\times10^{8}$ $M_{\odot}$ estimated from the $M_{\textrm{bulge}}-M_{\textrm{BH}}$ relation \citep{Bettoni03}. 

\subsection{Jet/Cavity Geometry}
\label{geometry}
\begin{figure}
\includegraphics[scale=0.24]{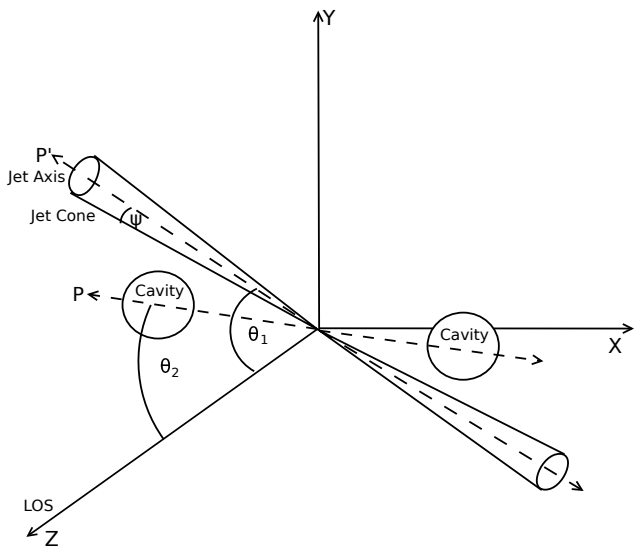}
\caption{The schematic plot of the jet/counterjet geometry. $\psi$ is the half jet cone angle,
	$\theta_1$ is the angle between the jet axis and the line-of-sight, and 
	$\theta_2$ is the angle between the cavity and the line-of-sight. P is the previous jet axis and P$^\prime$ is the
	current jet axis.
}
\label{cartoon}
\end{figure}

\begin{figure}
\includegraphics[scale=0.32,angle=-90]{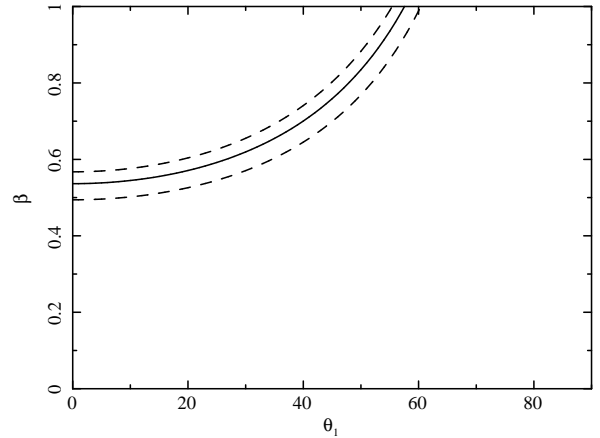}
\caption{The constraints on the jet speed and the angle to the line-of-sight for the jet to counterjet brightness ratio $R_{jet}$ of 20 (the solid line). Two dashed lines show the constraints for $R_{jet}$ of 15 and 25. Only the parameter space above the lines is allowed.}
\label{jetangle}
\end{figure}
Fig. \ref{cartoon} is a schematic diagram showing the geometry of the jets and cavities, viewed from the positive $z$-axis. 
The jets propagate out from the origin of the 3D axes into cones with half angle $\psi$. 
The angle between the jet and the line-of-sight (LOS) is $\theta_1$. Two symmetric cavities lie on each side with an angle of $\theta_2$ to
the LOS. The jet axis is assumed to be aligned with cavity previously (axis marked as P) and realigned to the current axis (marked as P$^\prime$).

Assuming that the jets are intrinsically symmetric, we can place constraints on jet velocity and the angle to the LOS
based on the jet to counterjet brightness ratio \citep[e.g.,][]{Giovannini94}:
\begin{equation}
	R_{\rm jet} = (1+\beta \cos\theta_{1})^{(2+\alpha)}(1-\beta \cos\theta_{1})^{-(2+\alpha)},
\end{equation}
where $R_{\rm jet}$ is the jet to counterjet brightness ratio, $\beta$ is the ratio of the jet velocity to the speed of light, 
$\theta_1$ is the angle to the LOS as shown in
Fig. \ref{cartoon}, and $\alpha$ is the spectral index. In Fig. \ref{jetangle}, we show the constraints on the intrinsic jet velocity and the angle between the jet and the LOS with a spectral index of 0.5 for 3C~88. The radio counterjet to the southwest is not detected. Based on the \vla\ 4.9 GHz image (Fig. \ref{radio} right), we put a lower limit of 20 for $R_{\rm jet}$, with the bright NE jet patch at $\sim$ 16 kpc from the nucleus and its counter position at the SW. As shown in Fig. \ref{jetangle}, the maximum value for $\theta_1$ is 57\arcdeg\ for $R_{\rm jet}=20$.

Can the narrow jet produce such a wide cavity as shown in Fig. \ref{image}? Here we list some possibilities. First, the light jets with relativistic cosmic rays (CRs) 
can create fat cavities. The CR-dominated jets are so light that they are quickly decelerated in the ICM due to their low inertia and momentum, then the CR pressure drives
the lateral jet expansion to form fat lobes, which are mainly formed by jet backflows and supported by the CR pressure \citep{Guo11}.
Second, the jet intermittency can also help. The recent study shows that multiple jets produce a single, large cavity system
if the reorientation angle between two successive jets is small and the time interval is sufficiently short so that they can
keep inflating the same cavity \citep{Cielo18}. This is consistent with the fact that the current radio jet is misaligned with the X-ray cavity in 3C~88. 
Another possibility to create fat bubbles is through ultrafast outflows (UFOs, $v>$ a few $10^{4}$ km\,s$^{-1}$) 
\citep[e.g.,][and references therein]{sternberg07,Soker16}. 
In general-relativistic, radiative magneto-hydrodynamic simulations \citep[GR-RMHD, e.g.,][]{Gaspari17}, the accretion energy is transformed
within 100s of gravitational radii into the UFOs, which become more massive at the kpc scale and decelerate, inflating the bubble
and thermalizing the kinetic energy mainly through turbulent mixing.

Previous studies \citep[e.g.,][]{Rafferty06,Hlavacek-Larrondo12} found that the shapes of cavities vary from almost perfectly circular to elongated.
\citet{Shin16} found that the shapes of non-circular cavities follow a bimodal distribution,
in that cavities are elongated either along the jet direction or perpendicular to the jet direction.
In a recent study, \citet{Guo15} showed that the shapes of young cavities are closely related to various jet properties, 
e.g., jet density, velocity, energy density, and duration. In general, very light internally subsonic jets produce bottom-wide cavities 
elongated along the perpendicular direction to the jet, while the heavier internally supersonic jets produce top-wide cavities
elongated along the jet direction. The long term evolution of cavity depends significantly on the viscosity of the hot ICM,
and on the amount of entrainment along the jet/outflow path from the pc to 10 kpc scale.
From Fig. \ref{image}, the eastern cavity in 3C~88 appears more elongated in the direction perpendicular to the jet 
(although projection effects may make the cavities appear more circular).
However, the cavity is located away from the jet axis and the angle between the jet axis and the radial direction 
from the AGN center to the cavity center is $\sim50\arcdeg$. This is suggested to be due to jet reorientation (see
section \ref{reorientation}), which makes it hard to infer the jet properties directly from the cavity shape.

The region of the eastern cavity seen in Fig. \ref{image} can be approximated as a circle with a radius of 23 kpc, projected at a distance 
of 28 kpc from the AGN.
Assuming that the cavity is a sphere in three-dimensional space, we can make a rough estimate of
the angle between the LOS and the cavity.
For an angle $\theta_2$ between the LOS and cavity, the real distance of the cavity to the center is therefore 28 kpc/sin($\theta_2)$.
We first simulated the 3D gas distribution with its temperature and density following the deprojected profiles. The cavities are assumed to be spheres
devoid of the gas. Then we projected the simulated data on the sky plane with a different angle $\theta_2$. We used
two circular regions with the same size, with one free of the cavity and the other overlaid with the projected cavity, and
calculated their surface brightness ratio. As expected, we found that the surface brightness ratio increases with the angle $\theta_2$, e.g.,
the surface brightness ratio increases by about $\sim50$\% as $\theta_2$ ranges from 30\arcdeg\ to 90\arcdeg. We can place constraints on the angle $\theta_2$
based on the surface brightness ratio, e.g., a threshold of 1.1 giving a lower limit on $\theta_2=32\arcdeg$.
Notice that in real observations, the surface brightness
contrast between the ambient region and the cavity region is more striking due to the bright surrounding rims or enhancements, 
with a much larger surface brightness ratio than in our simulations.

\subsection{Jet Reorientation}
\label{reorientation}
With deep X-ray observations, systems with multiple pairs of X-ray bubbles (or cavities) at different positional angles have been observed \citep[e.g.,][]{Bogdan14, Randall15}.
One scenario for the formation of multiple pairs of X-ray bubbles is jet reorientation model, in which the jet axis was initially along 
the direction of one pair of bubbles, and along a different direction later (the jet activity may switch off before changing directions).
The jet reorientation models include the conical precession, which reproduces well the observed radio structures in some radio sources \citep[e.g.,][]{Gower82,Cox91,Dunn06b}, 
and other realigning mechanisms, e.g., due to the binary black hole merger or the instabilities of accretion disc \citep[e.g.,][]{Dennett02, Gopal12}.
3C~88 provides a great case to study jet reorientation and two methods are discussed in this section.

\subsubsection{Method I: X-ray cavity + radio jet}
Although only a single pair of X-ray cavities was found in 3C~88 (in fact only the eastern cavity is pronounced), the eastern cavity is misaligned with the radio lobe. The position angle of the cavity, relative to the central AGN, also differs from that of the radio jet. 
The radio jet of 3C~88 was assumed to be previously lying in the direction from the nucleus to the cavity center, and currently aligned with the observed radio jet.
If both the X-ray cavity and the radio jet are in the plane of sky (or $\theta_1$ = $\theta_2$ = 90\arcdeg), the angle between the eastern cavity and the current radio jet is about 50\arcdeg.
If we assume this is the change of the angle between the previous and current jet axis during the span of eastern bubble rising, the jet reorientation rate is $\sim8\arcdeg$ per $10^{7}$ yr, 
which is between the precession rates found in Abell~3581 \citep{Canning13} and NGC~1275 \citep{Dunn06b}.
However, since the inclination angles (angle between the jet and the LOS or between the cavity and LOS) can be different from 90\arcdeg,
we need to estimate the projection effect on the estimated rate.
In Fig. \ref{cartoon}, if we view the system along the polar axis ($z$-axis), the polar angles for jet and cavity are $\theta_1$ and $\theta_2$,
respectively, and the separation between them in azimuth is $\delta A\sim50\arcdeg$. Therefore, the real angular separation $\delta\phi$ 
between them in 3D is given as:
\begin{equation}
	\cos(\delta\phi) = \cos(\theta_1)\cos(\theta_2)+\sin(\theta_1)\sin(\theta_2)\cos(\delta A),
\end{equation}
which can be expressed as:
\begin{equation}
	\sin^{2}(\delta\phi/2) = \sin^{2}(\delta\theta/2) + \sin(\theta_1)\sin(\theta_2)\sin^{2}(\delta A/2),
\end{equation}
where $\delta\theta=\theta_2-\theta_1$.
When $\delta\theta$ is small compared to $\delta A$ (e.g., $\theta_1\sim\theta_2$), we have  $\sin(\delta\phi/2)\simeq \sin(\theta)\sin(\delta A/2)$. Assuming the same inclination angle of 40\arcdeg\ for jet and cavity,  the real angle between the previous and current jet axis is 63\% of the separation angle between the cavity and radio jet on the plane of sky. Since the actual distance from the AGN to the center of cavity is now 56\% longer, the actual jet reorientation rate is only 40\% of the estimated value without accounting for the projection effect.
Generally with the constraints derived for $\theta_1$ and $\theta_2$ in the last section, we estimate $\delta\phi \sin(\theta_2)/50$, which measures the correction factor on the jet reorientation rate from the projection. This ratio ranges from $\sim$ 0.25 to 1.65, which can be applied on the baseline reorientation rate of $\sim8\arcdeg$ per $10^{7}$ yr estimated above.
\subsubsection{Method II: radio jet curvature}
In Fig. \ref{radio} the northeastern radio jet starts to bend at a projected distance of $\sim55$ kpc 
with a deflection angle of $\sim15\arcdeg$. Two hotspots at the northeast and at least one hotspot at the southwest are detected.
This type of radio morphology, also observed in other systems, (e.g., 3C~123, Cygnus A) can be explained by
jet reorientation \citep[e.g.][]{Cox91,Steenbrugge08,Pyrzas15}. 
In this scenario, the jet initially strikes the wall of 
the cocoon producing the jet curvature as it feeds the primary hotspot. The jet then interacts with the cocoon at a
very sharp angle and generates a new primary. The old primary is then observed as a secondary hotspot. In this picture,
the radio jet of 3C~88 was previously lying in the direction from the nucleus to the
secondary hotspot, and it is currently aligned with the observed radio jet. 

Based on the eastern bent jet we can make a rough estimation of the time-scale. As seen in the radio image,
the narrow radio jet maintains its width and does not widen even at a large radius ($\sim55$ kpc). We do not expect 
a significant jet deceleration due to the low ICM density and expected low stellar mass loss rates. 
Therefore, the jet remains relativistic until it is close to the hot spot. Assuming a constant speed of $\sim0.5c$ (see Fig. \ref{jetangle})
for jet travelling from the center to 55 kpc, and a lower speed from the bent point to the hotspots, we can estimate the 
travel time, which depends mostly on the latter speed. 
Assuming a jet speed of $0.01c$ or $0.05c$ \citep[e.g.,][]{Scheuer95,Carilli96} we obtained a travel time of
4.3 Myr or 1.2 Myr, corresponding to a jet reorientation rate of 4\arcdeg\ or 15\arcdeg\ per Myr.

We notice that the estimated jet reorientation rates from two methods differ a lot, since they are probing different epochs, e.g., method I probes the average rate whereas method II probes the instantaneous one.
The origin of jet reorientation is likely tied to the accretion history of the central SMBH. An intriguing scenario is related to the transition between FR I and FR II radio sources \citep[e.g.,][]{Garofalo2010}. On the other hand, in the framework of the CCA, warm/cold clouds and filaments collide inelastically, which can boost the accretion rate and suddenly alter the SMBH spin and thus the jet orientation \citep[e.g.,][]{Gaspari12a}.
In fact only CCA can likely drive such a large variability able to alter rapidly the BH spin: this is because of the random orientation of the raining clouds \citep[see also][]{King06,King08}. 
	Hot accretion is too smooth to rapidly change the spin (and is also spherically symmetric);  disc-driven jets have quite fixed orientation perpendicular to the disc and it 
	takes very long time to alter the disc morphology and fully change its direction as in a merger (rather than accreting different already inclined clouds as in CCA).

\subsection{Cavity Power in Radio Loud Groups}
\label{CavityPower}
\begin{figure}
\includegraphics[scale=0.55]{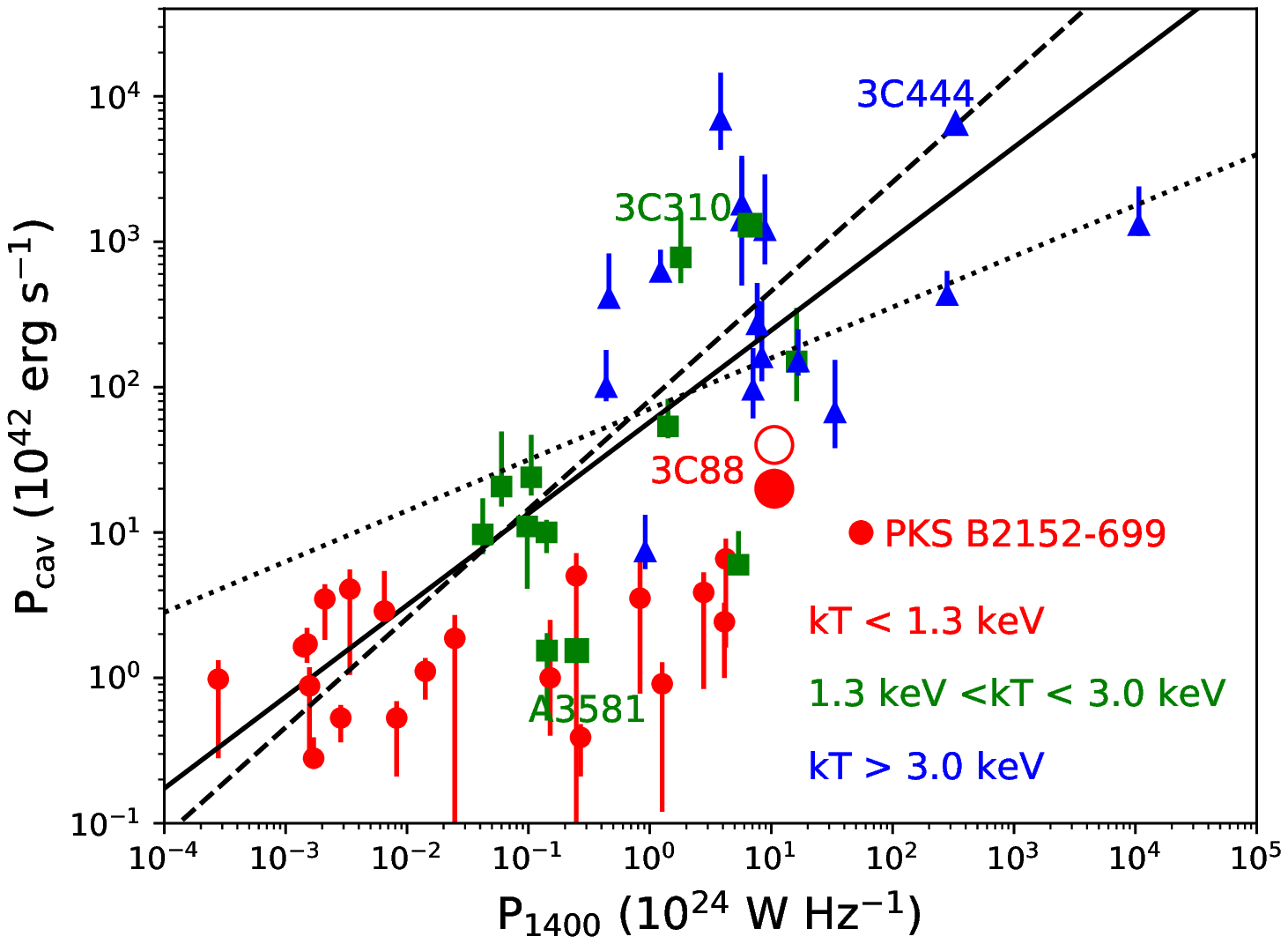}
\caption{Cavity power vs radio power at 1.4 GHz for galaxy groups and clusters (red: $kT<1.3$ keV, green: $1.3<kT<3.0$ keV, blue: $kT>3.0$ keV) from \citet{Birzan08}, \citet{Cavagnolo10}, and \citet{OSullivan11}. The dotted, dashed, and solid lines are the best-fitting relations from their work, respectively. For some repetitive systems, we used the values from \citet{OSullivan11}. The differences in cavity power are generally within a factor of 2 for all repetitive systems, except for NGC~6269, whose cavity power is about 7 times smaller in \citet{Cavagnolo10}. New studies on NGC~5813 found three pairs of X-ray cavities \citep{Randall11,Randall15}. However, the total cavity power is consistent with the value used in the plot. We included four additional systems, PKS~B2152-699 \citep[][the cavity power is estimated using the buoyant rise time as for other systems in the plot]{Worrall12}, 3C~310 \citep{Kraft12}, 3C~444 \citep{Croston11} and Abell~3581 \citep{Canning13} in the plot. The large solid red circle denotes the location of 3C~88, 
	 whose cavity power is among the largest in galaxy groups (please see Section 6.4 for discussion).
The empty red circle represents 3C~88 assuming there are a pair of cavities with the same power.
}
\label{pcav}
\end{figure}

X-ray cavities provide a direct estimate of the power output of the central AGN, from the cavity enthalpy (the minimum energy required to 
form a cavity) and the timescale over which the cavity has formed. The mechanical power of the cavities is found to be correlated with
the AGN radio luminosity, e.g., $P_{\textrm{radio}}-P_{\textrm{cav}}$ relation \citep{Birzan08,Cavagnolo10,OSullivan11}. In general, a large cavity power
is expected for luminous radio galaxies. However, as shown in Fig. \ref{pcav}, the cavity power found in radio-loud galaxy 
groups (e.g., $P_{1.4\textrm{GHz}}\sim1-10\times10^{24}$ W Hz$^{-1}$, and $kT<1.3$ keV) in the previously studied 
samples \citep[e.g.,][]{Birzan08,Cavagnolo10,OSullivan11} is less than $10^{43}$ erg s$^{-1}$.
For 3C~88 with $L_{1.4\textrm{GHz}}\sim10^{25}$ W Hz$^{-1}$, we measured the total mechanical power of cavities of $\sim2\times10^{43}$ erg s$^{-1}$,
which is at least two or three times higher than what is found in other radio luminous groups, except for PKS~B2152-699 \citep{Worrall12}.
The cavity power of PKS~B2152-699 is $\sim3\times10^{43}$ erg s$^{-1}$ using the dynamical time, or $\sim1\times10^{43}$ erg s$^{-1}$ using the buoyant rise time as for 3C~88. 
In either case, the cavity power of the two radio luminous galaxy groups, 3C~88 and PKS~B2152-699, are comparable. X-ray cavities in PKS~B2152-699 are over 2 times smaller than 3C~88's in radius as 3C~88's X-ray cavity is the largest known in galaxy groups.

If the cavity center does not lie in the plane of sky, we need to estimate the projection effect on the estimated cavity power.
Since the actual distance from the AGN to the center of cavity and the size of cavity would increase, the pressure at the
center of cavity would be lower, while the volume of cavity would increase, hence affecting the estimation of total enthalpy.
In addition, the estimated age of the cavity will increase. Assuming a low limit of $\sim30\arcdeg$ for inclination angle $\theta_2$,
we obtained a new cavity power of $5\times10^{43}$ erg s$^{-1}$.

Despite the uncertainties in estimating the cavity powers, e.g., the estimation of the cavity volume, a natural question is why
the measured cavity powers in groups are generally low, even for those radio luminous systems. \citet{Sun09} showed that groups
with a luminous cool core ($L_{0.5-2\textrm{keV}}\ge10^{41.7}$ erg s$^{-1}$) do not host luminous radio 
AGN ($L_{1.4\textrm{GHz}}>10^{24}$ W Hz$^{-1}$).
The groups with luminous radio AGN only have small coronae or small cool cores, so the extended X-ray atmosphere that encloses
the radio lobes is faint. As both cavities and shocks are easier to detect in high density regions, these features become subtle
in groups, especially for groups hosting strong radio AGN. The \chandra\ constraints on the cavity
power may only include the power deposited in cool cores and the estimated power should be taken, generally, as a lower limit.

\subsection{Shock Energy}
\label{sectionEs}
\begin{figure*}
\begin{tabular}{ll}
\includegraphics[scale=0.52]{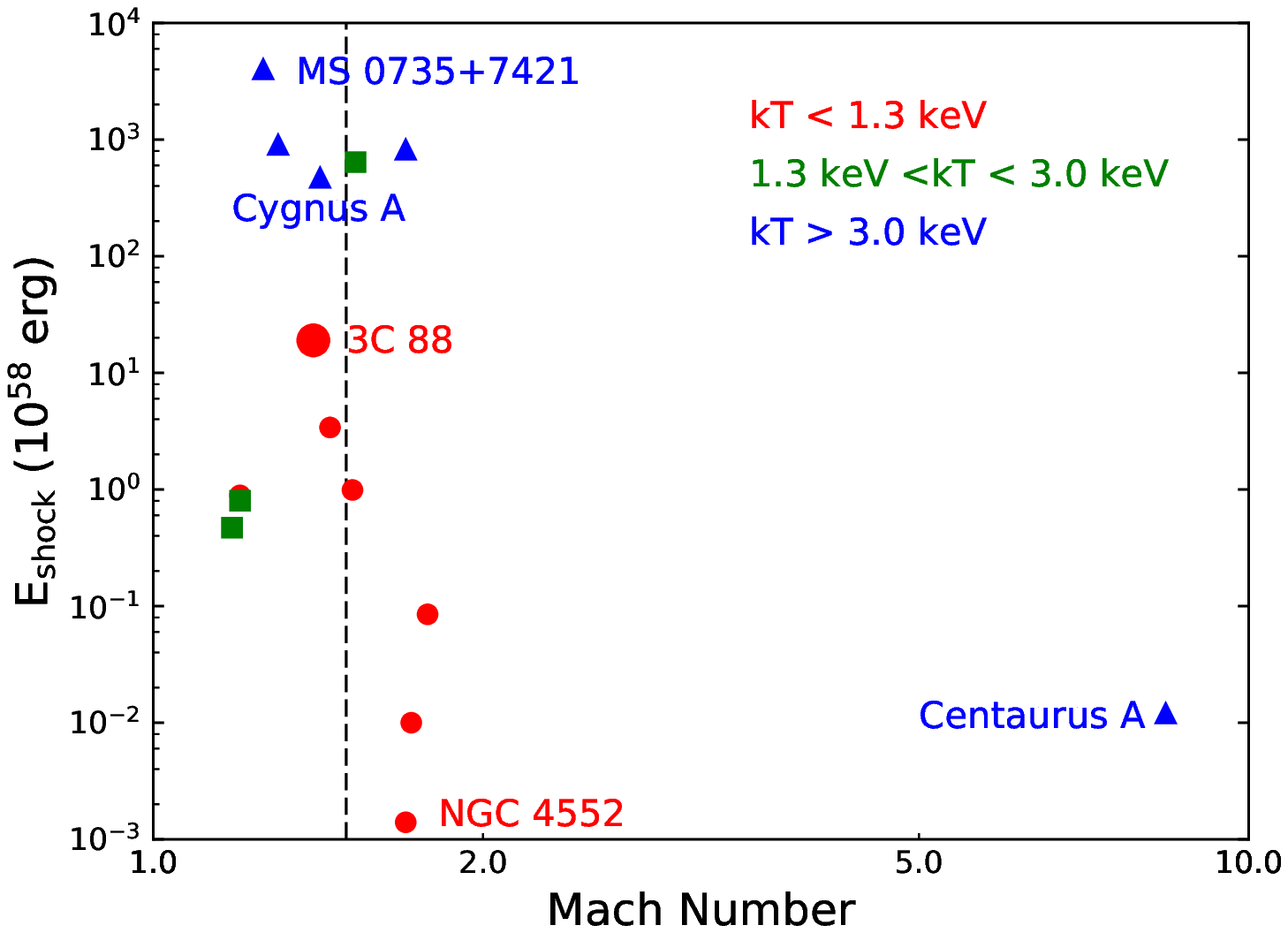}
&
\includegraphics[scale=0.52]{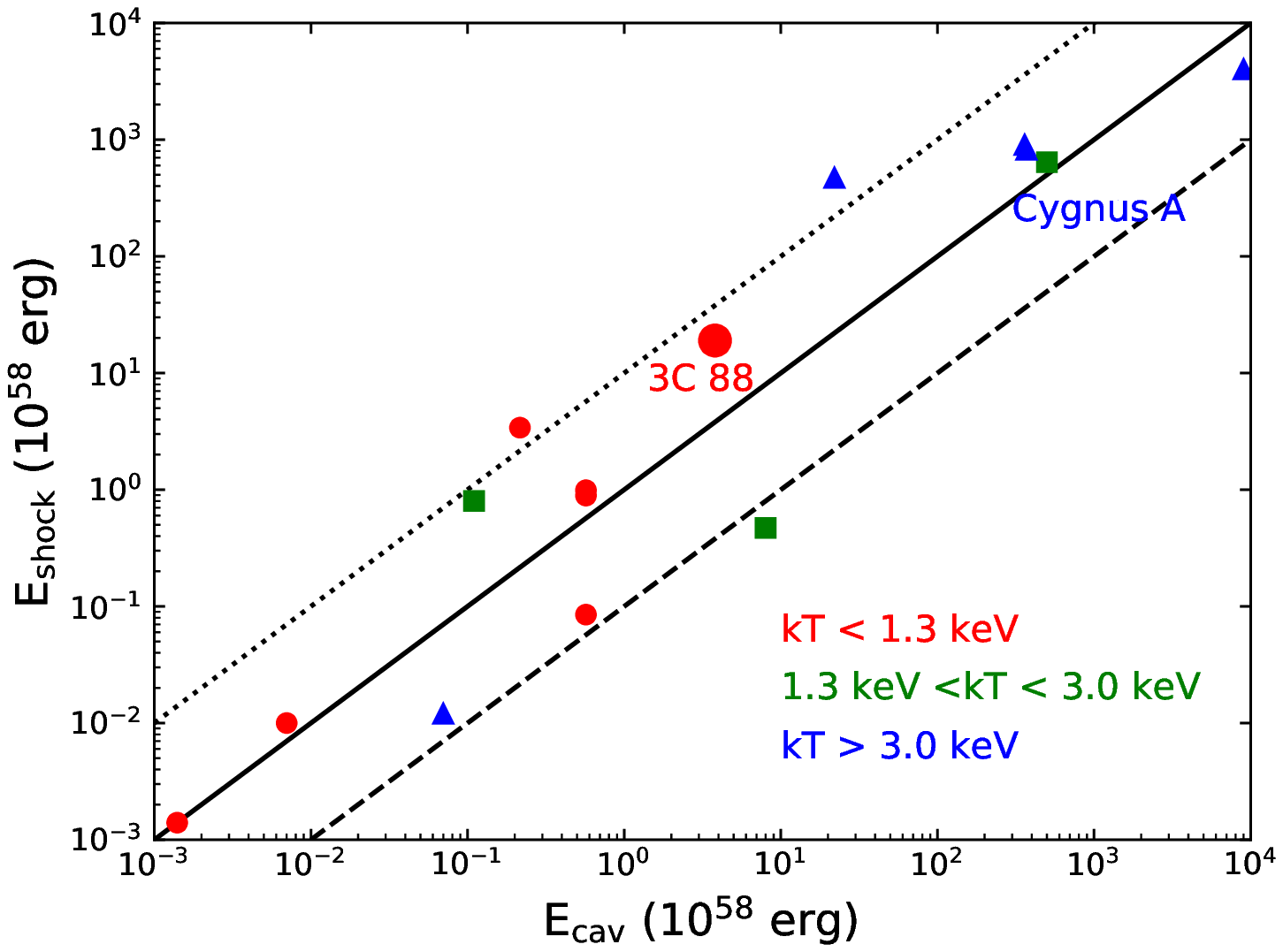}
\end{tabular}
\caption{Shock energy vs. Mach number (left) and shock energy vs. cavity enthalpy (right) for
	galaxy groups and clusters (red circle: $kT < 1.3$ keV, green square: $1.3 < kT < 3.0$ keV, and blue triangle: $kT > 3.0$ keV) from 
	the literature (see details in section \ref{sectionEs}). The dashed line on the left panel represents Mach number=1.5. The dotted, solid, and dashed lines 
	on the right panel represent $E_{shock}/E_{cav}=$ 10, 1, and 0.1, respectively. The large red circle denotes 3C~88.
}
\label{shock}
\end{figure*}
During an AGN outburst, the radio jets inflate the cavities in the ICM and drive shocks. 
The energy released by the AGN is partitioned between the total internal energy of the cavities and the work done
by them as they expand. The latter part drives the shock.
Following \citet{Randall11}, we estimated the shock energy from the pressure increase across
the shock front. Assuming a total energy $E$ is added to a gas of volume $V$, the pressure increase is roughly $\Delta P\sim E/V$.
For a shock with a known Mach number $M$, the ratio of the post- and pre-shock pressure $(P+\Delta P)/P$, the total
shock energy is therefore given as:
\begin{equation}
        E\approx P_1V(\frac{P_2}{P_1}-1) ,
	\label{equation}
\end{equation}
where $P_1$ and $P_2$ are the pre- and post-shock pressure respectively and their ratio depends on $M$.
The estimated shock energy from equation \ref{equation} is expected to modestly underestimate the total shock energy since the Mach number is expected
to be somewhat higher at earlier times. 
The shock energy estimated using equation \ref{equation} is found to agree reasonably well within a factor of a few 
with the shock energy calculated from the hydrodynamical central point explosion model based on the surface brightness \citep[e.g.,][]{Randall11}.

Using an ellipsoidal geometry for the volume contained within the shock front with semi-major and semi-minor axes of 85.7 kpc and 68.0 kpc 
(as shown in Fig. \ref{SB}) and the azimuthally averaged pressure profile from Fig. \ref{Tprofile}, we estimated a shock energy of 
$1.9\times10^{59}$ erg.
To make a comparison, we checked through the literature for the systems with available Mach numbers, shock energies, and cavity enthalpies,
e.g., Hydra~A \citep{Nulsen05b, Wise07}, MS~0735+7421 \citep{McNamara05,Vantyghem14}, Centaurus~A \citep{Croston09}, Cygnus~A \citep{Birzan04,Snios18}, 
3C~444 \citep{Croston11}, M87 \citep{Churazov01,Forman05,Forman07,Forman17}, Abell~2052 \citep{Blanton09}, 3C~310 \citep{Kraft12},
NGC~4552 \citep{Machacek06}, NGC~4636 \citep{Baldi09}, HCG~62 \citep{Gitti10}, 
NGC~5813 \citep[][notice NGC~5813 shows three distinct AGN outbursts.]{Randall11,Randall15}.
In Fig. \ref{shock} (left), we show the Mach number versus shock energy. 
The shock energies span almost 7 orders of magnitude, with a shock energy of $4\times10^{61}$ erg for MS~0735+7421 and $\sim10^{55}$ erg for NGC~4552,
while the Mach numbers range from $1-2$, except for Centaurus A. The shock energy in 3C~88 is the largest among the low temperature groups. 
Although it has a similar Mach number and shock radius to 3C~88, Cygnus~A, by far the nearest powerful radio galaxy, has a shock energy about 20 times larger
than that of 3C~88.
In Fig. \ref{shock} (right), we show the cavity enthalpy versus the shock energy.
Either shock energy or cavity enthalpy can be the most significant form of 
mechanical energy produced by AGN outbursts, with their ratio generally varying within $0.1-10.0$.
In 3C~88, the ratio of the shock energy to the total cavity enthalpy is about 5.0.

\subsection{Bright Rims and Arms}
\label{rimsarms}
\begin{figure}
\includegraphics[scale=0.29]{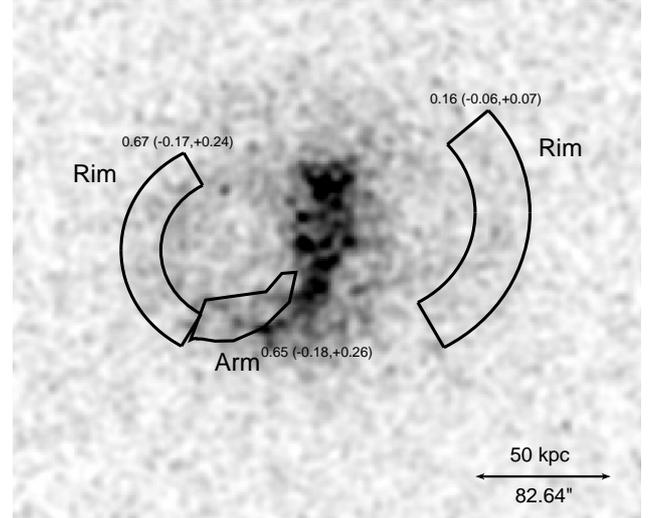}
	\caption{0.7-2.0 keV \chandra\ image of 3C~88 to show the regions and the measured abundances that are discussed in section \ref{rimsarms}.
}
\label{rims}
\end{figure}
The abundance map shown in Fig. \ref{TZ_map} suggests enhanced abundance along the eastern bright rim and arm. Several interesting regions with enhanced X-ray surface brightness are defined in Fig. \ref{rims}.
There are two possible explanations for the presence of the enhanced X-ray features. One explanation is the rims could be shells
of gas shocked during the supersonic 
expansion of the bubbles \citep[e.g.,][]{Fabian03, Forman07}. Alternatively, if the bubbles are rising in a subsonic phase of expansion,
the dense and cool gas from the cluster center could be lifted by the rising bubbles
(as in, e.g., M87: \citeauthor{Churazov01} 2001, \citeauthor{Simionescu08} 2008, \citeauthor{Werner10} 2010; NGC~1399: \citeauthor{Su17} 2017; NGC~4472: \citeauthor{Gendron17} 2017; 
Hydra~A: \citeauthor{Kirkpatrick09} 2009, \citeauthor{Gitti11} 2011).

The spectra for each region, from all three observations, were then fitted simultaneously with the model
{\tt TBABS*APEC} with fixed absorbing column density. We found temperatures of $1.36^{+0.10}_{-0.03}$ and $1.33^{+0.21}_{-0.07}$ 
keV for the eastern and western rims, respectively, which are higher than those from adjacent regions. The metallicity in the western
rim is $0.16^{+0.07}_{-0.06}$ solar, while it is $0.67^{+0.24}_{-0.17}$ in the eastern rim. 
The metallicity enhancement surrounding the eastern cavity implies that the hot, metal rich gas was lifted by the rising X-ray bubble
from the group center \citep[e.g.,][]{Kirkpatrick15}.
The relation between the radius of the eastern rim (taken as the ``iron radius'', the maximum radius at which a significant enhancement in 
metallicity has been detected) and the cavity enthalpy is generally consistent with that found in a previous study \citep{Kirkpatrick15}.
The pronounced cavity and high abundance in the eastern rim show that the gas remains unmixed with the nearby material. 
On the other hand, the less significant western cavity and the relatively 
low abundance in the western rim suggest the gas has been well-mixed with the surrounding gas.

In addition to the bright rims, we also observed an enhanced spiral-arm-like feature in the south in Fig. \ref{rims}. We performed a similar spectral analysis for the arm as for the rims and found the projected temperature of $1.03\pm{0.02}$ keV and metallicity of $0.65^{+0.26}_{-0.18}$.
The arm is connected with the eastern rim. Its relatively high abundance suggests it also has not well mixed with the surrounding material. It should be noted that the measured abundances in this section are projected values and the actual abundance enhancement with the eastern rim and arm should be stronger.

\section{Summary}
We have analyzed 105 ksec \chandra\ observation of the radio-loud galaxy group 3C~88 and presented the properties of the hot gas related to AGN feedback. Our results are summarized as follows:
\begin{itemize}
	\item The eastern X-ray cavity is very prominent with a diameter of $\sim$ 50 kpc and is centered $\sim28$ kpc away from the nucleus. The enthalpy of the eastern cavity is $3.8\times10^{58}$ erg. Both the cavity size and enthalpy are the largest known in groups. We estimate that the cavity age is about 60 Myr. The power of the eastern cavity is $\sim2.0\times10^{43}$ erg s$^{-1}$, among the largest known in galaxy groups.
	\item From the surface brightness discontinuity, we infer a Mach $\sim1.4$ shock front. The estimated shock energy is $1.9\times10^{59}$ erg, which is about five times larger than the cavity enthalpy. 
	\item The rest-frame 2 - 10 keV luminosity of the central AGN is $\sim 4\times10^{41}$ erg s$^{-1}$. The X-ray spectrum shows a moderate intrinsic absorption with a column density of $\sim 2\times10^{22}$ cm$^{-2}$. 
	\item The eastern cavity is not well matched spatially with the radio lobe and is not aligned with the jet axis, which suggests jet reorientation. 
		We used different methods to estimate the average and instantaneous jet reorientation rates.
	\item The rim and arm features surrounding the X-ray cavities with enhanced X-ray surface brightness show metallicity enhancement, 
		suggesting the gas was lifted by the rising X-ray bubbles from the group center.
	\item Warm gas is detected in the central region, extending to at least $\sim5$ kpc, and from a compact knot about 16 kpc northeast of the nucleus with potentially more diffuse features. The multiphase gas state, low minimum condensation ratios, rapid jet precession, and large mechanical power are consistent with the scenario of CCA 
			boosting mechanical AGN feedback.

		There is no reason to believe 3C~88 is special as strong radio AGN exist in many low-mass halos and jets can reorient. This detailed study of 3C~88 also suggests that galaxy groups with powerful radio AGN can indeed have large cavity power. As groups have shallower potential, smaller core size and less hot gas than clusters \citep[e.g.][]{Sun12}, X-ray cavities in groups can easily extend beyond the small, dense core. Deep \chandra\ observations are then required to reveal the subtle features in groups. On the other hand, these heating events also leave imprints on the X-ray properties of groups over time \citep[e.g.][]{McNamara12,Sun12}.

\end{itemize}
\section{Acknowledgements}
Support for this work was provided by the National Aeronautics and Space Administration through \chandra\ Award Number GO4-15115X, GO5-16115X, GO5-16146X, GO7-18121X,
and AR7-18016X issued by the \chandra\ X-ray Center, which is operated by the Smithsonian Astrophysical Observatory for and on behalf of 
the National Aeronautics Space Administration under contract NAS8-03060. We also acknowledge the support from NASA/EPSCoR grant 
NNX15AK29A and NSF grant 1714764. M.G. is supported by NASA through Einstein Postdoctoral Fellowship Award Number PF5-160137 issued by the \chandra\ X-ray 
Observatory Center, which is operated by the SAO for and on behalf of NASA under contract NAS8-03060.
Basic research in radio astronomy at the Naval Research Laboratory is supported by 6.1 Base funding.
This research has made use of data and/or software provided by the High Energy Astrophysics Science Archive Research 
Center (HEASARC), which is a service of the Astrophysics Science Division at NASA/GSFC and the High Energy Astrophysics Division 
of the Smithsonian Astrophysical Observatory.
The NRAO \vla\ Archive Survey image was produced as part of the NRAO \vla\ Archive Survey, (c) AUI/NRAO.
The National Radio Astronomy Observatory is a facility of the National Science Foundation operated under cooperative agreement by Associated Universities, Inc.
We thank the staff of the \gmrt\ that made these observations possible. \gmrt\ is run by the National Centre for Radio Astrophysics of 
the Tata Institute of Fundamental Research. 
We thank Megan Donahue, Thomas Connor, Rachel Salmon and Dana Koeppe for a spectroscopic observation of 3C~88 with SOAR early 2018.
This research has made use of data based on observations obtained with the Apache Point Observatory 3.5-meter telescope, which is owned and operated by the Astrophysical Research Consortium. APO observations were supported by the F. H. Levinson Fund of the Silicon Valley Community Foundation.
\bibliographystyle{mnras}
\bibliography{my}

\appendix

\section[]{}
\label{appendix}
\begin{figure}
\begin{center}
\hbox{
	\includegraphics[scale=0.5]{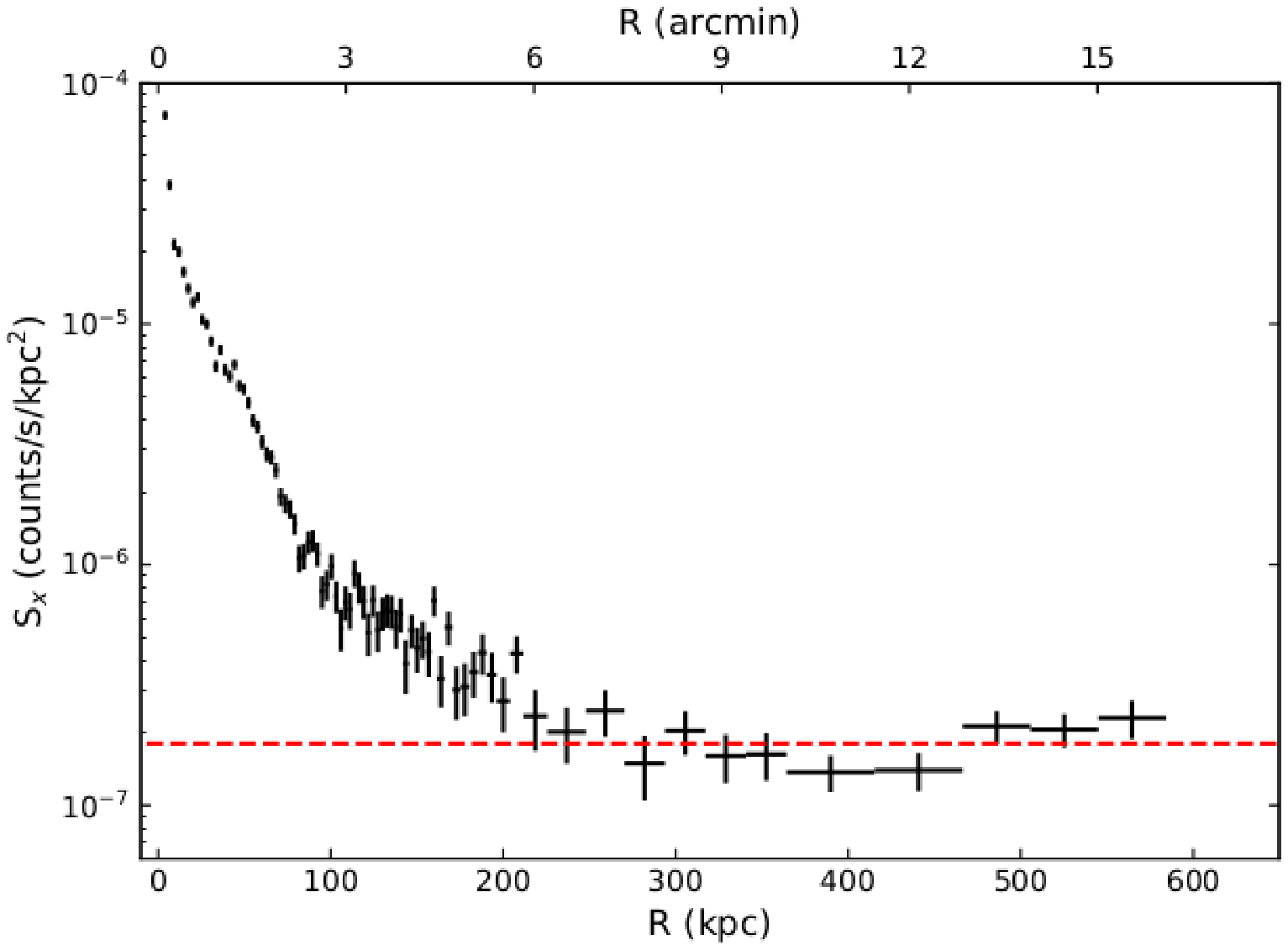}
}
\caption{\label{sbbkg}The surface brightness profile of galaxy group 3C~88 in the 0.7-2.0 keV band. The red dashed line
represents the surface brightness of the local X-ray background.}
\end{center}
\end{figure}

Fig. \ref{sbbkg} shows the surface brightness profile of galaxy group 3C~88 in the 0.7-2.0 keV band. The surface
brightness is approximately constant beyond $\sim8$\arcmin\ from the center of 3C~88. 

\end{document}